\RequirePackage{rotating}
\documentclass[format=acmsmall,review=false,screen=false]{acmart}

\pdfoutput=1

\usepackage{booktabs} 
\usepackage[english]{babel}
\usepackage[caption=false,font=footnotesize]{subfig}
\usepackage{amsmath}
\usepackage{amsfonts,amssymb}
\usepackage{mathtools}
\usepackage{caption}
\usepackage{url}
\usepackage{tabularx}
\usepackage[flushleft]{threeparttable}
\usepackage{xspace}
\usepackage{array}
\usepackage{adjustbox}
\usepackage{multirow}
\usepackage{bm}
\usepackage{paralist}
\usepackage{balance}
\usepackage{pifont}
\usepackage{relsize}
\usepackage{float}
\usepackage{pdflscape}
\usepackage{lscape}
\usepackage[b]{esvect}

\usepackage{tikz}
\usetikzlibrary{angles}
\usetikzlibrary{quotes}
\usetikzlibrary{calc}
\usetikzlibrary{intersections}
\usetikzlibrary{fit}
\usetikzlibrary{shapes.geometric}
\usetikzlibrary{datavisualization}
\usetikzlibrary{datavisualization.formats.functions}	
\usetikzlibrary{shadings}
\usetikzlibrary{external}
\tikzsetexternalprefix{tikz/}

\usepackage{pgfplots}
\pgfplotsset{compat=1.13}

\graphicspath{{./imgs/}}

\makeatletter
\def\url@urlfontstyle{%
 \@ifundefined{selectfont}{\def\UrlFont{\sf}}{\def\UrlFont{\bf\ttfamily}}}
\makeatother
\usepackage{ifthen}
\newboolean{authnotes}

\setboolean{authnotes}{false}

\ifthenelse{\boolean{authnotes}}
{
\newcommand{\si}[1]{\footnote{{\bf Silvia:} #1}}
\newcommand{\lm}[1]{\footnote{{\bf Luca:} #1}}
\newcommand{\mz}[1]{\footnote{{\bf Marco:} #1}}
\newcommand{\ch}[1]{\footnote{\color{blue}{\bf Carsten:} #1}}
}
{
\newcommand{\si}[1]{}
\newcommand{\lm}[1]{}
\newcommand{\mz}[1]{}
\newcommand{\ch}[1]{}
}

\newcommand\figref[1]{Fig.~\ref{#1}}
\newcommand\tabref[1]{Table~\ref{#1}}
\newcommand\secref[1]{Sec.~\ref{#1}}

\newcommand\question[1]{\begin{quote} \emph{#1} \end{quote}\vspace{1mm}}

\newcommand{\eg}{\emph{e.g.},\xspace}

\newcommand{\ie}{\emph{i.e.},\xspace}
\newcommand{\etc}{etc.\xspace}
\newcommand{\capt}[1]{\mdseries{\emph{#1}}}

\newcommand{\fakepar}[1]{\vspace{1mm}\noindent\textbf{#1.}}

\newcommand{\examplePaper}[1]{\vspace{1mm}\noindent\textbf{Example $\rightarrow$ #1:}}
\newcommand{\furtherExamples}{\vspace{1mm}\noindent\textbf{Completing the picture.}\xspace}

\setdefaultleftmargin{\parindent}{}{}{}{}{}
\newcommand{\st}{synchronous transmissions\xspace}
\newcommand{\lt}{link-based transmissions\xspace}
\newcommand{\ieee}{IEEE 802.15.4\xspace}
\newcommand{\ble}{BLE\xspace}
\newcommand{\wifi}{Wi-Fi\xspace}
\newcommand{\cps}{CPS\xspace}
\newcommand{\iot}{IoT\xspace}

\newcommand{\pthree}{P$^3$\xspace}

\newcommand{\dsss}{DSSS\xspace}
\newcommand{\prr}{PRR\xspace}
\newcommand{\csma}{CSMA\xspace}

\newcommand{\meter}{\ensuremath{\,\text{m}}\xspace}

\newcommand{\us}{\ensuremath{\,\mu\text{s}}\xspace}

\newcommand{\kHz}{\ensuremath{\,\text{kHz}}\xspace}

\newcommand{\GHz}{\ensuremath{\,\text{GHz}}\xspace}
\newcommand{\dB}{\ensuremath{\,\text{dB}}\xspace}

\newcommand{\ppm}{\ensuremath{\,\text{ppm}}\xspace}

\newcommand{\percent}{\ensuremath{\,\text{\%}}\xspace}

\newcommand{\kmax}{\ensuremath{\,\text{k}_{max}}\xspace}

\newfloat{eqfloat}{tbh}{aux}

\setcopyright{none}

\begin{document}
\title[Synchronous Transmissions in Low-Power Wireless]{Synchronous Transmissions in Low-Power Wireless:\\ A Survey of Communication Protocols and Network Services}
\author{Marco Zimmerling}
\orcid{0000-0003-1450-2506}
\affiliation{%
  \institution{TU Dresden}
  \department{Center for Advancing Electronics Dresden (CfAED), Networked Embedded Systems Lab}
  \streetaddress{Helmholtzstrasse 18, BAR IV56 (D wing, fourth floor)}
  \city{Dresden}
  \postcode{01069}
  \country{Germany}}
\email{marco.zimmerling@tu-dresden.de}
\author{Luca Mottola}
\orcid{0000-0003-4560-9541}
\affiliation{%
  \institution{Politecnico di Milano}
  \department{Dipartimento di Elettronica, Informazione e Bioingegneria, Networked Embedded Software Lab}
  \streetaddress{Via Golgi 42, Building 22 (room 319, third floor)}
  \city{Milan}
  \postcode{20133}
  \country{Italy}}
\affiliation{%
  \institution{RISE SICS}
  \department{Computer Systems Laboratory, Networked Embedded Systems Group}
  \streetaddress{Electrum, Isafjordsgatan 22/Kistagången 16 (sixth floor)}
  \city{Kista}
  \postcode{16440}
  \country{Sweden}}
\email{luca.mottola@polimi.it}
\author{Silvia Santini}
\orcid{0000-0002-0882-2004}
\affiliation{%
  \institution{Universit\'a della Svizzera italiana (USI)}
  \department{Faculty of Informatics}
  \streetaddress{Via Buffi 13, Informatics Building (office 108, level 1)}
  \city{Lugano}
  \postcode{6900}
  \country{Switzerland}}
\email{silvia.santini@usi.ch}


\begin{abstract}
Low-power wireless communication is a central building block of Cyber-physical Systems and the Internet of Things.
Conventional low-power wireless protocols make avoiding packet collisions a cornerstone design choice.
The concept of synchronous transmissions challenges this view.
As collisions are not necessarily destructive, under specific circumstances, commodity low-power wireless radios are often able to receive useful information even in the presence of superimposed signals from different transmitters.
We survey the growing number of protocols that exploit synchronous transmissions for higher robustness and efficiency as well as unprecedented functionality and versatility compared to conventional designs.
The illustration of protocols based on \st is cast in a conceptional framework we establish, with the goal of highlighting differences and similarities among the proposed solutions.
We conclude the paper with a discussion on open research questions in this field.
\end{abstract}




\maketitle



\section{Introduction}

Cyber-physical Systems (\cps) and the Internet of Things (\iot) blur the boundary between the physical and cyber domains,  enabling powerful applications in areas as diverse as industrial automation, smart health, and ambient intelligence~\cite{lee08cps,stankovic14iot}.
Their concrete realization relies on embedded hardware and software with distinctive features.
Crucially, to enable pervasive and untethered deployments, the devices communicate wirelessly and are powered by batteries or small capacitors that store limited amounts of energy, possibly harvested from the environment~\cite{bhatti16energy}.


Fundamental design trade-offs are pursued to stay within the small energy budgets.
Low-complexity physical layers enable low-power wireless transceiver implementations, but this comes at the cost of low data rates (tens of kbit/s to a few Mbit/s) and limited communication ranges (tens of meters to a few kilometers) depending on the technology.
\ieee and Bluetooth Low Energy (\ble) are examples of short-range low-power wireless technologies, whereas LoRa and Sigfox belong to the corresponding class of long-range technologies.
In addition, the radios and other device components are duty cycled by switching them on and off based on application requirements and communication demands to reduce energy consumption; however, a device cannot receive any packet while its radio is off.

The goal of minimizing energy consumption is thus at odds with other key performance objectives, such as increasing the fraction of packets delivered to the destinations across the unreliable wireless channel, which impacts \emph{reliability}, or reducing the time needed to do so, which affects \emph{latency}.
Because of limited communication ranges, multi-hop communication is often employed whereby intermediate nodes relay packets from sources to destinations, which further compounds the problem.
Since the early 2000's, the tension among those conflicting goals has motivated a profusion of research efforts in low-power wireless communication and networking~\cite{karaki04routing,demirkol06mac}.

\begin{figure}[t]
 \centering
 \subfloat[Using link-based transmissions nodes A, B, and C transmit one after another to avoid collisions at the receiver R.\label{fig:link_based_transmissions}]{\includegraphics[width=0.55\textwidth]{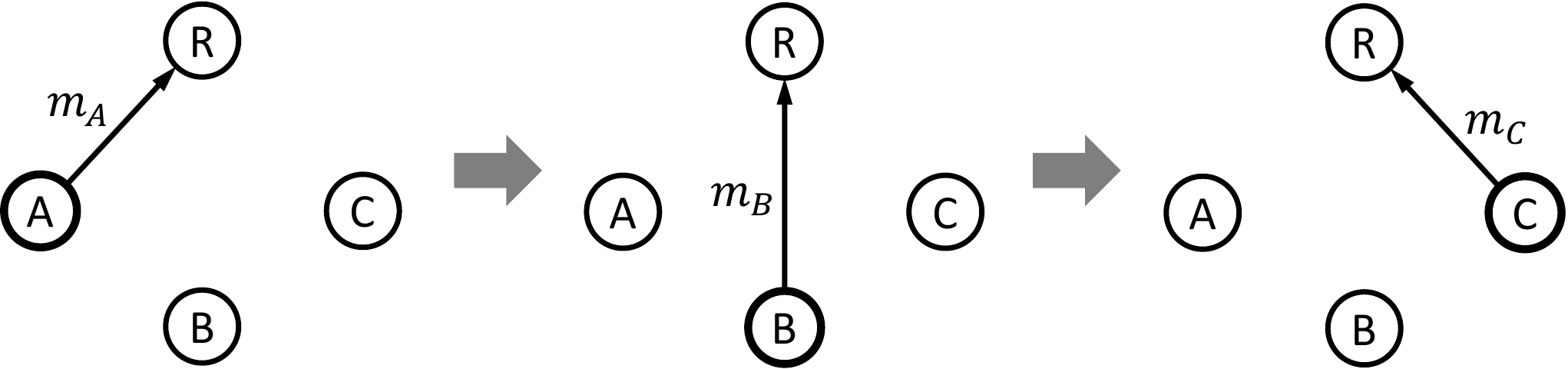}}
 \quad
 \subfloat[Using synchronous transmissions all nodes transmit at the same time.\label{fig:synchronous_transmissions}]{\includegraphics[width=0.35\textwidth]{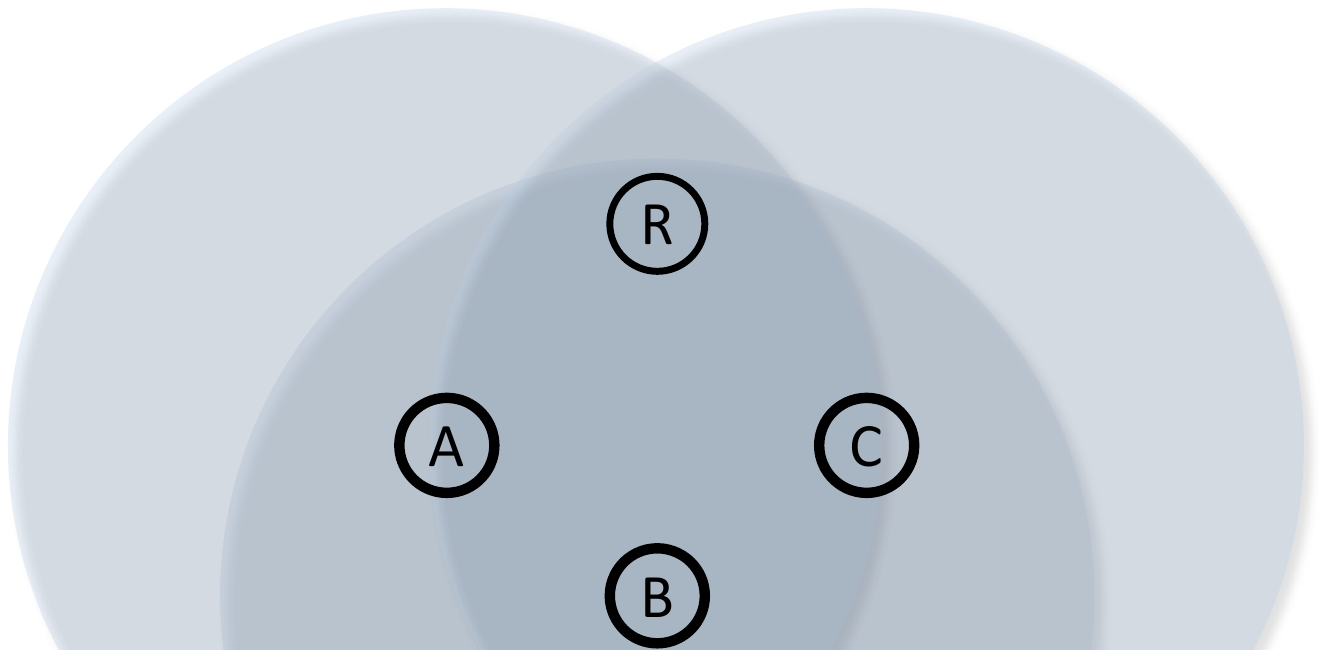}}
 \caption{Illustration of link-based transmissions in (a) versus synchronous transmissions in (b).}
 \label{fig:link_based_vs_synchronous_transmissions}
\end{figure}

\fakepar{Link-based and synchronous transmissions} Low-power wireless communication is broadcast in nature: the radio waves emitted by a transmitter travel through the air and can be received by any device in communication range and with the appropriate equipment.
If a device is in the communication range of multiple transmitters, that device sees the superposition of multiple signals whenever two or more transmitters transmit at the same time on the same carrier frequency.
The common belief suggests that such \emph{packet collisions} are destructive and prevent a device from receiving any useful information.
To achieve high reliability and energy efficiency, conventional low-power wireless protocols try to avoid collisions using techniques such as carrier sensing, handshaking, and scheduling.
As a result, neighboring devices transmit packets one after another over \emph{point-to-point links}, as illustrated in \figref{fig:link_based_transmissions}.
We refer to this concept as \emph{\lt}, which are used by the vast majority of low-power wireless protocols, for example, as it allows to adopt well-known approaches from wired networking such as routing.

Since 2010, a new breed of low-power wireless communication protocols emerged based on the concept of \emph{\st}.
Unlike \lt, the tenet of \st is that collisions are not necessarily destructive.
Rather, by purposely letting multiple nodes transmit at the same time on the same carrier frequency, as shown in \figref{fig:synchronous_transmissions}, with high probability a node can receive useful information nonetheless.
This allows protocols using synchronous transmission to embrace the broadcast nature of low-power wireless communication instead of hiding it under the artificial abstraction of point-to-point links.

There exist related concepts in  wireless communications, including cooperative diversity~\cite{sendonaris2003user1,sendonaris2003user2,laneman2004cooperative} and non-orthogonal multiple access~\cite{saito2013systemlevel} based on, for example, successive interference cancellation~\cite{verdu1998multiuser}.
While \st can be considered a specific variant of those concepts, a key difference is that \st are applicable to off-the-shelf commodity low-power wireless devices and existing physical layers, including \ieee, \ble, and LoRa.
Thus, millions of cheap microcontroller-based devices deployed around the world support \st out of the box.

The gains and benefits of \st over \lt range from orders of magnitude lower latencies~\cite{du16pando} and energy consumption~\cite{istomin16crystal} to providing the ability functionality such as virtual synchrony~\cite{ferrari13virtual} that was previously thought unfeasible~\cite{stankovic03realtime}.
Researchers from academia and industry also built on the concept of \st when participating to international scientific competitions~\cite{schuss17competition}, demonstrating, for example, the robustness of this technique even under severe interference from co-located networks.
Synchronous transmissions thereby offer protocol designers a radically different perspective and unprecedented possibilities.
Nevertheless, compared to traditional protocols built on \lt, the use of \st is often based on different assumptions, requires modified protocol layering and architectures, and demands alternative approaches to evaluating performance.

\begin{figure}[tb]
\begin{center}
\includegraphics[scale=.7]{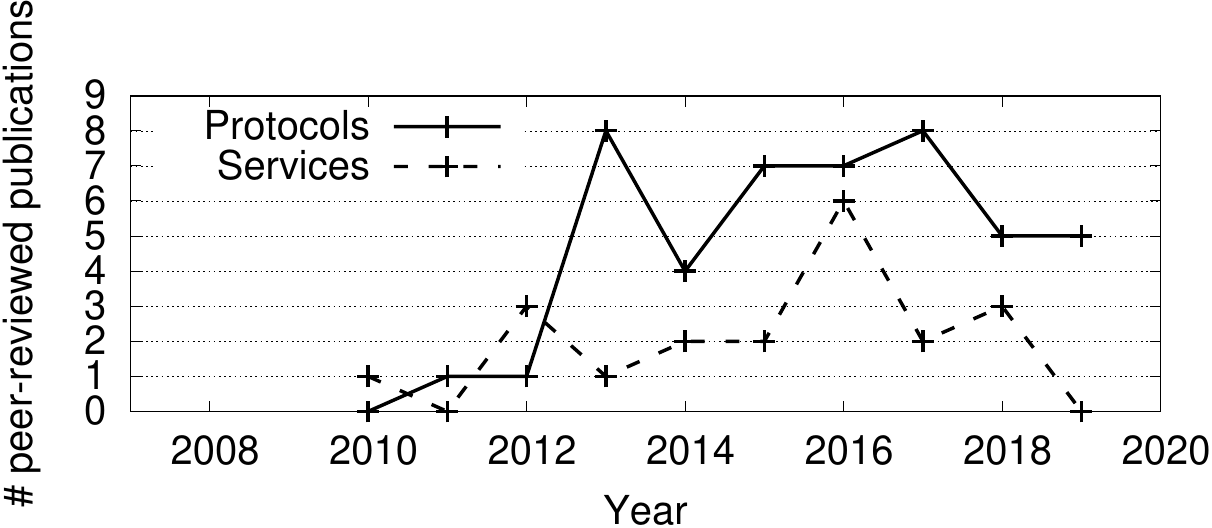}
\caption{Number of peer-reviewed conference and journal publications over time as surveyed in this article, covering the design of communication protocols and network services based on \st.}
  \label{fig:papersOverTime}
\end{center}
\end{figure}

\figref{fig:papersOverTime} shows the take-up of \st over time for the design of communication protocols and network services in the peer-reviewed publications we include in this survey, demonstrating the growing interest.
Following pioneering work that appeared between 2010 and 2012, the bulk of the technical and scientific contributions is arguably found in the last five to six years.
As the body of published material grows, it is now time to consolidate the knowledge and to reflect on what are the challenges and opportunities yet to be explored.
A recent tutorial~\cite{chang2019constructive} makes a first attempt toward this goal, but covers only a small fraction of the published material and does not present a conceptual framework that elicits the similarities and differences among existing solutions.

\fakepar{Contribution and road-map} This paper presents an extensive analysis of the use of \st in the design of low-power wireless communication protocols and network services.
Coherently surveying the topic requires the necessary background for readers to build upon.
\secref{sec:primer} provides a primer on synchronous transmissions in low-power wireless communications, focusing on the conditions and effects that enable a successful packet reception in the presence of collisions and the resulting benefits.
Readers interested in a more detailed treatment of the underlying physical-layer principles are referred to our complementary paper~\cite{firstpart}, which also surveys works seeking to understand and improve the reliability of the synchronous transmission technique itself.

\secref{sec:comm} represents the core of this paper.
It contributes an extensive analysis and classification of existing low-power wireless communication protocols employing \st.
The semantics we associate to this term defines the scope of our work: In low-power wireless networks, a \emph{communication protocol} is regarded as providing applications with functionality for data collection, data dissemination, or peer-to-peer interactions~\cite{Levis:2004:ENA:1251175.1251176,hui2008ip,minakov2016comparative,mottola2017makesense,Dutta:2005:SST:1129601.1129732,Kothari:2008:DSM:1371607.1372740,mottola2012middleware,javed2018internet}.
Therefore, a communication protocol enables the exchange of \emph{user-defined data} among physical devices, therefore offering an even minimal programming interface that allows \emph{applications} to access its functionality.
Our review of existing works in this area unfolds through the analysis of key dimensions that characterize the state of the art, providing a novel conceptual framework that serves as guidance through the literature.
For each dimension of this framework, we first discuss the existing individual instances and then describe an existing protocol as a paradigmatic example.


There exists also a number of solutions that use \st to realize network primitives not expressly meant to transport application data, but to provide higher-level network functionality (\eg consensus) or auxiliary network services (\eg time synchronization).
We discuss those solutions in \secref{sec:serv}.
Then, in \secref{sec:agenda}, we outline a variety of open challenges and research questions.
These issues pertain to fields as diverse as understanding and exploiting the peculiar features of \st in the design of communication protocols, conceiving novel network stack designs to better leverage \st, the alternative definition of communication patterns and network services, the integration of low-power wireless systems using \st with standard network infrastructures, and standardization processes.
Many of these challenges require expertise from diverse fields that hitherto experience little cross-fertilization.
This paper is also the opportunity to foster such an exchange of knowledge and insights between the relevant disciplines.
We end the paper in \secref{sec:end} with brief concluding remarks.



\section{A Primer on Synchronous Transmissions in Low-power Wireless}
\label{sec:primer}

This section provides a brief overview of the physical-layer principles that enable synchronous transmissions on off-the-shelf low-power wireless platforms, and works out the key advantages of synchronous transmissions especially from a protocol design perspective.

\subsection{Receiving in the Presence of Packet Collisions}

Understanding under what circumstances a node can successfully receive a packet in the presence of collisions motivated a wealth of research~\cite{wilhelm14reception,liao16revisiting}.
Current knowledge indicates that three effects play an important role, often in combination, depending on the receiver implementation and on the properties of the incoming signals: capture effect~\cite{leentvaar76capture}, message-in-message effect~\cite{manweiler09order}, and constructive interference~\cite{ferrari11efficient}.
We briefly discuss these physical-layer effects in the following, focusing on the most relevant aspects needed as background for the remainder of this article.
For a more comprehensive description we refer the reader to our complementary paper~\cite{firstpart}.

\fakepar{Capture and message-in-message effects}
The capture effect enables a receiver to correctly receive one of the synchronously transmitted packets if the incoming signals satisfy certain power and timing conditions.
Specifically, the difference in received power between one signal and the sum of all other signals plus noise must exceed the
\emph{capture threshold}.
If such strongest signal arrives first, the receiver will correctly receive it with high probability, while treating all other signals as noise.
Otherwise, if such strongest signal arrives after a weaker signal, it must do so within the \emph{capture window}, that is, before the receiver finishes receiving the preamble of the earlier packet.
The reason is that, after receiving the first valid preamble, low-power wireless radios suspend preamble search until they have received the complete packet.
Thus, the radio cannot ``switch'' to a stronger signal if it arrives too late, resulting in an erroneous reception as the stronger signal destroys the weaker signal.

The capture threshold and capture window depend on the physical layer.
For example, using \ieee radios with O-QPSK modulation in the 2.4\GHz band, experimental studies report a capture threshold of 2-3\dB~\cite{dutta10design,liao16revisiting,koenig16maintaining} and a capture window of 128\us~\cite{yuan13talk}.
The capture effect is a general phenomenon already exploited in a variety of other popular wireless technologies, such as \ble~\cite{roest15}, Bluetooth~\cite{cordeiro03}, IEEE 802.11 variants~\cite{lee07}, LoRa~\cite{bor16}, and sub-GHz radios~\cite{whitehouse05}.

The message-in-message effect essentially plays like the capture effect, but without a timing condition.
It may occur with more advanced radios, for example, certain \wifi transceivers, that also perform preamble search after the reception of a valid preamble~\cite{manweiler09order}.
Such radios can ``abort'' an ongoing reception and ``switch'' to another signal that exceeds the capture threshold, possibly also multiple times.
Eventually the strongest of all overlapping signals is received with high probability.
Because it does not matter when the strongest signal arrives, the probability of correctly receiving a message despite a collision is higher than with radios that only feature the capture effect.

\fakepar{Constructive interference}
If all signals arrive with similar power, a successful reception is only possible if the transmitted packets are identical and so-called \emph{constructive interference} occurs~\cite{ferrari11efficient}.
True constructive interference means that the signals perfectly overlap, that is, there are no time and phase offsets.
However, even if transmitters can compensate for different path delays, and thus make the signals arrive without time and phase offsets at the receiver, the phase offsets would vary \emph{throughout} the reception because of different carrier frequencies across the transmitters.

Carrier frequency offsets are inevitable without a shared clock: the radios' oscillators generating the carriers run asynchronously at different frequencies within a certain tolerance range.
For example, according to the \ieee standard, the oscillators are allowed to drift by up to $\pm$40\ppm, which translates into a maximum carrier frequency offset of 192\kHz between sender and receiver for the 2.4\GHz band.
Thus, in a real network of commodity low-power wireless devices, true constructive interference is highly improbable.
Rather, studies reporting higher median and variance in received power suggest that, as a result of the carrier frequency offsets, there are alternating periods of constructive and destructive interference during a reception~\cite{dutta08backcast,koenig16maintaining,liao16revisiting,wang15disco}.

Periods of destructive interference can cause bit errors at the receiver, and without additional measures those lead to an erroneous packet  reception.
One possible measure is a channel code that enables a receiver to correct a certain number of bit errors.
For example, the direct-sequence spread spectrum~(\dsss) technique used by 2.4\GHz \ieee radios tolerates twelve incorrect bits (called chips) per symbol.
With this error-correction capability, a receiver is able to successfully receive an entire packet despite a range of possible carrier frequency offsets if the time offset among the signals is less than 0.5\us~\cite{liao16revisiting}.
The ability to cope with carrier frequency offsets is strongly related to the receiver implementation, especially whether the demodulator works in a coherent or non-coherent fashion (\eg O-QPSK or MSK in case of 2.4 GHz IEEE 802.15.4)~\cite{escobar2019improving}.
This may explain empirical results indicating that synchronous transmissions work even in the absence of a suitable error-correcting mechanism~\cite{alnahas2019concurrent}, such as most \ble variants or certain sub-GHz radios.

In summary, although the term constructive interference is technically incorrect and somewhat misleading, it nonetheless describes the fact that it \emph{appears} to a link-layer protocol as if the transmissions interfere constructively, whereas the underlying carrier signals do not.
The impact of physical-layer parameters, receiver implementations, and error-correction mechanisms on the occurrence of constructive interference is an important topic for future research (see \secref{sec:agenda}).

\subsection{Benefits of Synchronous Transmissions}

As discussed, synchronous transmissions only work under certain conditions.
It is not obvious whether it is easy or difficult to meet those conditions, and if doing so can improve the communication performance and reliability compared to link-based transmissions.
To illustrate this, let us assume that the link from node A to the receiver R in \figref{fig:link_based_transmissions} is of excellent quality, that is, it has a packet reception ratio~(\prr) very close to 100\percent when using link-based transmissions.
In this case, it is extremely unlikely that the \prr at the receiver R improves when nodes B and C transmit synchronously with node A.
It is much more likely that the \prr with synchronous transmissions slightly degrades because a certain chance exists that none of the above physical-layer effects occurs~\cite{liao16revisiting}, and a packet collision rather happens that prevents any packet from being correctly received.
Although real-world experiments show that synchronous transmissions improve \prr for low- and medium-quality links~\cite{wang13triggercast},
this alone does not explain the increasing popularity of synchronous transmissions in low-power wireless systems.

The works reviewed in this and our complementary paper~\cite{firstpart} demonstrate that creating the conditions under which synchronous transmissions work in real-world low-power power wireless networks is possible with no or only little overhead in a wide range of realistic settings.
Synchronous transmissions are thereby demonstrated to provide key benefits over link-based transmissions:
\begin{enumerate}
 \item \emph{Sender diversity.} Compared to link-based transmissions, the presence of multiple transmitters allows a receiver to benefit from \emph{sender diversity}. If the transmitters are at least one carrier wavelength apart (\eg about 0.125\meter at 2.4\GHz), their wireless links to the receiver are uncorrelated with high probability~\cite{tse2005fundamentals}. It is therefore unlikely that all links suffer fading or other adverse effects at the same time. This increases the chances of successful reception at a common receiver, compared to the expected reliability of link-based transmissions in comparable conditions and unless the pair-wise links are already near-perfect.
 \item \emph{Simplicity.} The observation that packet collisions are not necessarily harmful questions the need for information and mechanisms to avoid them. This includes the collection of information about the network state, such as a node's neighbors and the corresponding link qualities, as well as mechanisms like link estimators, link schedulers, and carrier sense multiple access~(\csma). These and other mechanisms are used extensively by conventional low-power wireless protocols to enable efficient and reliable multi-hop communication. The opportunity to drop them enables dramatically simpler protocol designs.
\end{enumerate}

These are the only benefits provided by the concept of synchronous transmissions \emph{itself}.
They are, however, fundamental in that they offer protocol designers a radically different perspective and unprecedented possibilities compared to link-based transmissions.
The extent to which these possibilities and benefits are exploited and possibly combined with other higher-level mechanisms depends on the (multi-hop) \emph{protocol design}; existing solutions show that the potential end-to-end gains are diverse and enormous.
For example, the degree of sender diversity is a function of the node density and the number of active synchronous transmitters, and can be combined with other forms of diversity including receiver diversity, obtained in settings with more than one receiver, frequency diversity, when nodes use channel hopping, and temporal diversity, when a message is transmitted multiple times and possibly by different sets of synchronous transmitters.

Based on this, multi-hop communication protocols have been designed whose logic is independent of the network state and whose performance can be accurately modeled by simple discrete-time Markov chains~\cite{ferrari12lwb,zimmerling13modeling}.
On the other hand, nothing prevents protocol designers from leveraging long-lived network state for optimizing performance in specific scenarios, possibly at the cost of reduced robustness to network dynamics.
The following two sections analyze the many ways in which synchronous transmissions have been used in the design of low-power wireless communication protocols and network~services.  


\section{Synchronous Transmissions in Communication Protocols}
\label{sec:comm}

\begin{figure}[tb]
\begin{center}
\includegraphics[scale=.3]{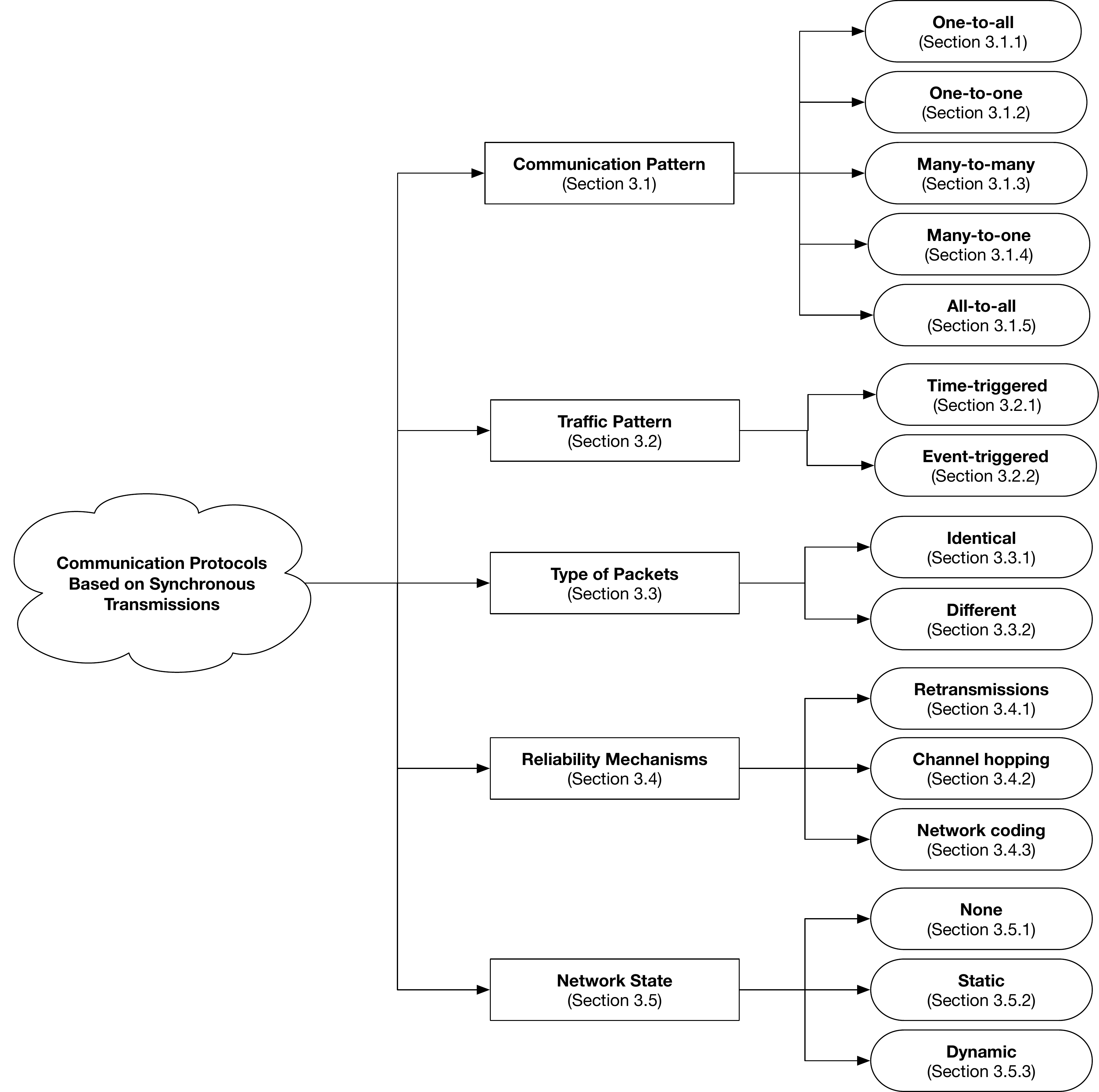}
\caption{Taxonomy and road-map for classifying communication protocols built on \st.}
  \label{fig:taxonomy_structure}
\end{center}
\end{figure}

This section analyzes existing multi-hop communication protocols built on \st.
To this end, we first identify key protocol characteristics allowing us to separate out the fundamental features of a protocol's design, independent of its concrete implementation. 
\figref{fig:taxonomy_structure} depicts the taxonomy resulting from this analysis, which also serves as road-map and reference throughout this section.
For every dimension of the taxonomy, we describe one protocol in more detail to exemplify a given feature, and briefly comment on other protocols whenever these build on top of the primary example or exhibit further interesting characteristics.
\tabref{tab:comm-overview} lists all protocols surveyed in this section and indicates their classification according to taxonomy in \figref{fig:taxonomy_structure}.

Existing protocols using \st are extremely diverse in terms of performance goals, design features, and implementation.
This diversity is also often reflected in different evaluation methodologies used in the original papers, ranging from, for example, simulations and table-top experiments to large-scale experiments on public testbeds.
It is therefore impossible to \emph{quantitatively} compare existing communication protocols in a fair way based on performance results reported in the original papers.
For this reason, we focus on providing a \emph{qualitative} comparison, and identify rigorous benchmarking as an important direction for future work (see \secref{sec:agenda}).




\addtolength{\tabcolsep}{-3pt}
\begin{sidewaystable}
 \centering
 \caption{Overview of synchronous transmission based communication protocols surveyed in \secref{sec:comm}, classified according to our taxonomy in \figref{fig:taxonomy_structure}. Protocols marked in bold are those described as the representative example of the corresponding dimension in our taxonomy. We abbreviate \emph{end-to-end} by \emph{e2e}.}
 \begin{small}
 \begin{tabular}{|p{3cm}|p{3.6cm}|p{2.3cm}|p{2.2cm}|p{6.1cm}|p{2.0cm}|}
  \hline
  \textbf{Protocol}                         & \textbf{Communication pattern} & \textbf{Traffic pattern} & \textbf{Type of packets} & \textbf{Reliability mechanisms} & \textbf{Network state}\\ \hline
  Glossy~\cite{ferrari11efficient}                        & \textbf{One-to-all}            & Time-triggered           & Identical                     & Retransmissions (local)       & None \\
  SCIF~\cite{wang12exploiting, wang13exploiting}          & One-to-all                     & Time-triggered           & Identical                     & Retransmissions (local)       & \textbf{Static} \\
  Splash~\cite{doddavenkatappa13splash}                   & One-to-all                     & Time-triggered           & Identical                     & \textbf{Channel hopping}, retransmissions (local, e2e), network coding & Static \\
  Ripple~\cite{yuan15ripple}                              & One-to-all                     & Time-triggered           & Identical                     & Network coding, channel hopping & Dynamic \\
  Zippy~\cite{sutton15zippy}                              & One-to-all                     & \textbf{Event-triggered} & Identical                     & Retransmissions (local)       & None \\
  Pando~\cite{du15pipelines,du16pando}                    & One-to-all                     & Time-triggered           & Identical                     & \textbf{Network coding}, channel hopping & Static \\
  RedFixHop~\cite{escobar16redfixhop}                     & One-to-all                     & Time-triggered           & Identical                     & Retransmissions (local)       & None \\
  LiM~\cite{zhang2017less}                                & One-to-all                     & Time-triggered           & Identical                     & Retransmissions (local)       & Dynamic \\
  BlueFlood~\cite{alnahas2019concurrent}                  & One-to-all                     & Time-triggered           & Identical                     & Channel hopping, retransmissions (local) & \textbf{None} \\
  CXFS~\cite{carlson13cxfs}                               & \textbf{One-to-one}            & Time-triggered           & Identical                     & Retransmissions (local)       & Dynamic \\
  Sparkle~\cite{yuan14sparkle}                            & One-to-one                     & Time-triggered           & Identical                     & Retransmissions (local)       & Dynamic \\
  PEASST~\cite{jeong14peasst}                             & One-to-one                     & Time-triggered           & \textbf{Identical}            & Retransmissions (local)       & Dynamic \\
  \pthree~\cite{doddavenkatappa14p3}                      & One-to-one                     & Time-triggered           & Identical                     & Channel hopping, retransmissions (e2e) & \textbf{Dynamic} \\
  RFT~\cite{zhang15rft}                                   & One-to-one                     & Time-triggered           & Identical                     & Retransmissions (local)       & Dynamic \\
  LaneFlood~\cite{brachmann16laneflood}                   & One-to-one                     & \textbf{Time-triggered}  & Identical                     & Retransmissions (local)       & Dynamic \\
  LWB~\cite{ferrari12lwb}                                 & \textbf{Many-to-many}          & Time-triggered           & Identical                     & Retransmissions (local)       & None \\
  eLWB~\cite{sutton2017thedesign}                         & Many-to-many                   & Event-triggered          & Identical                     & Retransmissions (local)       & None \\
  BLITZ~\cite{sutton2017blitz,sutton2019blitz}            & Many-to-many                   & Event-triggered          & Identical                     & Retransmissions (local)       & None \\
  WTSP \cite{suzuki17wireless}                            & Many-to-many                   & Time-triggered           & Identical                     & Retransmissions (local, e2e)    & None \\
  Blink~\cite{zimmerling17adaptive}                       & Many-to-many                   & Time-triggered           & Identical                     & Retransmissions (local)       & None \\
  Codecast~\cite{mohammad2018codecast}                    & Many-to-many                   & Time-triggered           & Different                     & Network coding, retransmissions (local) & Dynamic \\
  Mixer~\cite{mager17mixer,herrmann2018mixer}             & Many-to-many                   & Time-triggered           & \textbf{Different}            & Network coding, retransmissions (local) & Dynamic \\
  ByteCast~\cite{saha2017design}                          & Many-to-many                   & Time-triggered           & Different                     & Retransmissions (local)       & Dynamic \\
  Crystal~\cite{istomin16crystal,istomin2018interference} & \textbf{Many-to-one}           & Time-triggered           & Different                     & Channel hopping, retransmissions (local, e2e) & Dynamic \\
  Zimmerling et al.~\cite{zimmerling13modeling}           & Many-to-one                    & Time-triggered           & Identical                     & \textbf{Retransmissions} (local, e2e) & None \\
  Choco~\cite{suzuki13low}                                & Many-to-one                    & Time-triggered           & Identical                     & Retransmissions (local, e2e)    & None \\
  Sleeping Beauty~\cite{sarkar2016sleeping}               & Many-to-one                    & Time-triggered           & Identical                     & Retransmissions (local)       & Dynamic \\   
  Chaos~\cite{landsiedel13chaos}                          & \textbf{All-to-all}            & Time-triggered           & Different                     & Retransmissions (local)       & None \\
  \hline
 \end{tabular}
 \end{small}
 \label{tab:comm-overview}
\end{sidewaystable}
\addtolength{\tabcolsep}{3pt}

\subsection{Communication Pattern}
\label{sec:intent}

We begin by considering the communication patterns supported by existing protocols.
With this we refer to the \emph{intent} of the protocol, that is, the communication patterns offered to higher-level protocols or applications, and not to the protocol implementation.
We distinguish protocols that
\begin{itemize}
 \item support exactly \emph{one} sender;
 \item support any number of senders between one and all, referred to as \emph{many};
 \item support only scenarios in which \emph{all} nodes in the network operate as senders.
\end{itemize}

The number of receivers, on the other hand, indicates the number of distinct physical devices that the protocol \emph{intends} to deliver data to.
We distinguish protocols that transmit data to
\begin{itemize}
 \item exactly \emph{one} node in the network;
 \item any number of nodes between one and all, referred to as \emph{many};
 \item \emph{all} nodes in the network.
\end{itemize}


The above distinction results in nine different classes.
However, as we shall see below, concrete instances of protocols found in the literature only fall into a subset of these classes.

\subsubsection{One-to-all}
\label{sec:onetoall}

Approaches that enable a sender to transmit packets to all other nodes in the network belong to this class;
relevant examples include network flooding protocols that send single packets and data dissemination protocols that send multiple packets forming a larger data object. 

\examplePaper{Glossy} An example of a \emph{one-to-all} communication protocol that uses \st is Glossy~\cite{ferrari11efficient}.
Glossy provides network flooding and optionally time synchronization.
The work has played a seminal role in the evolution of research in this area as many communication protocols and network services build on or take inspiration from it. 

The single sender in Glossy is called \emph{initiator}.
It starts the flooding process by transmitting the packet, while all other nodes in the network have their radio in receive mode.
Provided propagation delays are negligible, the one-hop neighbors of the initiator receive the packet all at the same time.
Glossy then makes these nodes retransmit the received packet so that the retransmission delay is constant and equal for all involved devices.
This way, the one-hop neighbors of the initiator that received the packet retransmit it at nearly the same time.
Because these nodes are all sending the same packet, constructive interference can occur in the way described in \secref{sec:primer}.
This enables the two-hop neighbors of the initiator to receive and retransmit the packet, and so on.

To ensure a constant retransmission delay, Glossy executes a minimal, deterministic sequence of operations between packet reception and packet retransmission.
State-of-the-art radios offer features allowing to completely avoid this software processing. 
Further, Glossy floods are confined to reserved time slots, where no other tasks (application, operating system, \etc) are allowed to execute on a node.
This avoids interference on shared resources, which would cause unpredictable jitter.
Due to its operation, Glossy also enables nodes to time-synchronize with the initiator.

Without retransmission delay, Glossy achieves the theoretical lower latency bound for one-to-all communication of a single packet in multi-hop networks using half-duplex radios.
By retransmitting the packet a configurable number of times during a flood, Glossy exploits different forms of diversity for high reliability.
The diversity also de-correlates packet receptions and losses~\cite{zimmerling13modeling,Karschau2018}, which simplifies modeling and the design of higher-level functionality such as feedback controllers~\cite{mager2019}.
Nodes do not maintain any network state, which makes Glossy highly robust to network dynamics resulting from moving or failing nodes, interference from co-located wireless networks, \etc

\furtherExamples The Spine Constructive Interference-based Flooding (SCIF) protocol \cite{wang12exploiting, wang13exploiting} extends Glossy with a topology-control mechanism that makes it scale to high-density or large-scale networks, where the authors report that the performance of \st degrades.
RedFixHop~\cite{escobar16redfixhop} uses the same transmission technique as Glossy, but encodes the payload in the Medium Access Control (MAC) packet sequence number.
Upon the reception of a packet, a node forwards it using hardware-generated acknowledgments.
Because of the deterministic operation of the radio hardware, the retransmission delay is further reduced and better aligned across senders.
Less is More~(LiM)~\cite{zhang2017less} uses reinforcement learning to reduce the number of transmissions during a Glossy flood for improved energy efficiency, while maintaining high flooding reliability. 
BlueFlood~\cite{alnahas2019concurrent} demonstrates one-to-all communication using \st over BLE's physical layer.
Unlike Glossy and taking inspiration from~\cite{lim2017competition}, a node in BlueFlood performs $N$ back-to-back \st after receiving a packet for increased efficiency, and switches channel after each transmission and reception for increased reliability.    


While Glossy floods a single packet, a number of dissemination protocols have been designed to distribute multiple packets forming a larger data object.
The protocols mainly differ in the way they use pipelined Glossy floods over multiple channels with different forms of packet coding to disseminate all packets quickly and reliably.
Splash~\cite{doddavenkatappa13splash} uses XOR coding and Pando~\cite{du15pipelines,du16pando} uses Fountain coding, but both protocols assign channels to nodes based on their hop distance from the sender.
Ripple~\cite{yuan15ripple}, instead, assigns channels to packets and uses an erasure code.
These protocols rely on information about the network topology (\eg for channel assignment), which makes them suitable for scenarios in which the nodes are stationary throughout the dissemination process.






\subsubsection{One-to-one}
\label{subsec:one-to-one}
One-to-one protocols enable communication between a specific sender-receiver pair in a multi-hop network.
This type of communication primitive is beneficial if the information contained in a message is relevant for---or should be received by---only one destination node.

\examplePaper{CXFS} One of the first synchronous transmission based protocols that specifically target one-to-one communication is Concurrent Transmissions with Forwarder Selection (CXFS)~\cite{carlson13cxfs}.
Before the data exchange starts, CXFS performs a two-way handshake between the sender and the intended receiver.
During this handshake, the nodes determine whether they are forwarders on a shortest path between the source and the destination.
These nodes are included in the \emph{forwarder set} and remain active to relay messages between the sender and the receiver, whereas all other nodes switch their radios off to save energy.
Using a tunable parameter, the \emph{boundary} $b$, the forwarder set can be enlarged to make the data exchange more reliable at the cost of higher energy consumption.


While in Glossy a packet is retransmitted immediately upon reception, CXFS uses a dedicated schedule to organize packet transmissions.
In particular, every one-to-one data exchange occurs in time \emph{slots} of fixed length and a given number of \emph{frames} are included in each time slot.
At the beginning of the first frame in a slot, the node assigned to that slot can initiate the transmission of a packet.
Nodes receiving this packet will retransmit it at the beginning of the next frame, and so on.
This way of implementing \st is possible if the time synchronization among nodes is sufficiently accurate to guarantee a precise alignment of frames and slots.

\furtherExamples LaneFlood~\cite{brachmann16laneflood} is similar to CXFS, but operates directly on top of Glossy and lets nodes join the forwarder set with a given probability (refer to \secref{sec:time_triggered} for a more detailed description).
In a similar vein, the Practical Packet Pipeline (\pthree) protocol~\cite{doddavenkatappa14p3} supports the one-to-one transmission of multiple messages (\ie bulk traffic) along node-disjoint paths determined at a base station, based on global network-topology information collected beforehand (see \secref{sec:tree} for details).
Sparkle~\cite{yuan14sparkle} also targets one-to-one communication by carefully selecting subsets of nodes that participate in Glossy-like floods.
It uses a dedicated transmission-power and topology-control mechanism that seeks to boost the performance of \st.
The Reinforcement (RFT) protocol~\cite{zhang15rft} builds upon Sparkle to further improve reliability and energy efficiency, using a time-division multiple access (TDMA) approach for scheduling individual messages.



\subsubsection{Many-to-many}
\label{subsubsec:many-to-many}

Protocols in this category are capable of disseminating messages from any subset of nodes in the network (senders) to any other desired subset of nodes (receivers).
Many-to-many communication is the most universal pattern and essential for many applications and system services, including data replication (\eg for increased fault tolerance~\cite{schneider90}), certain programming abstractions~\cite{mottola11}, and distributed control tasks (\eg to coordinate a swarm of aerial drones~\cite{preiss17}).


\begin{figure}[tb]
\begin{center}
\includegraphics[width=.7\linewidth]{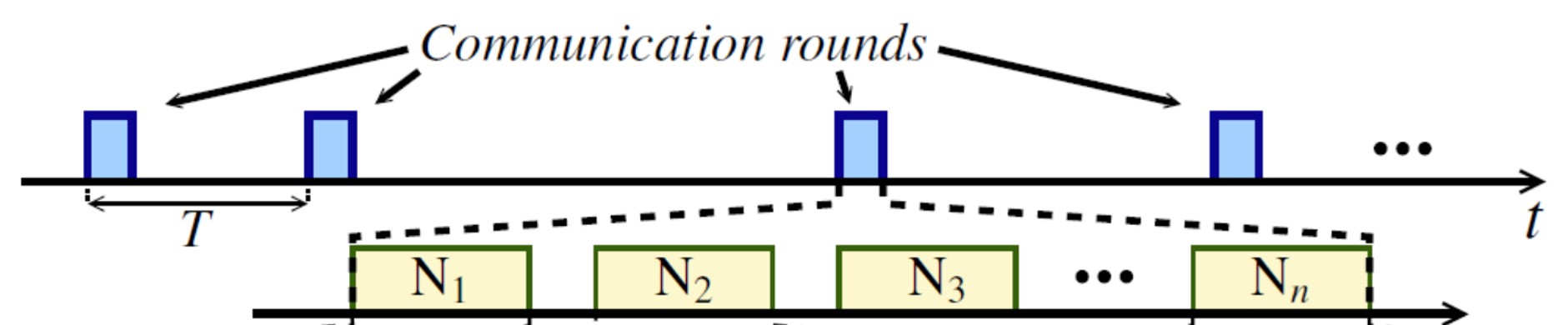}
\caption{Many-to-many communication in LWB~(taken from~\cite{ferrari12lwb}). \capt{LWB structures communication into rounds. A round contains a series of slots, $\textup{N}_{1}, \ldots, \textup{N}_{n}$, each of which serves to transmit one message using a Glossy flood.}}
  \label{fig:LWB}
\end{center}
\end{figure}

\examplePaper{LWB} The Low-power Wireless Bus (LWB)~\cite{ferrari12lwb} provides to a higher-layer protocol and the application the abstraction of a shared bus, wherein each node can potentially receive all transmitted messages.
As shown in \figref{fig:LWB}, LWB organizes communication into \emph{rounds}, which occur with a variable \emph{period} $T$.
A communication round contains up to $n$ \emph{slots}.
Every slot serves to transmit one message from a sender to all other nodes in the network using a Glossy flood; messages are then filtered at the receiver side and passed to a higher-layer protocol or the application only if the node is one of the intended receivers, as specified in the message header.
That is, LWB realizes many-to-many communication through a series of one-to-all transmissions followed by message filtering.
If the number of messages to be sent exceeds the maximum number of slots $n$ in a round, the many-to-many communication extends across multiple consecutive rounds.

When a node intends to send or receive messages, it first needs to notify LWB's central coordinator, called \emph{host}.
To this end, a node sends a \emph{stream request} in a \emph{contention slot}, which is a dedicated slot that is not assigned to a specific sender.
For prospective senders the stream request contains, among others, the inter-packet interval with which the stream generates new packets.
A node can request multiple streams and dynamically modify or remove a stream by sending a corresponding request to the host.
At the end of each round, the host computes the communication schedule for the next round based on all active streams.  
The host transmits this schedule twice for increased reliability: in the last slot of the current round and in the first slot of the next round.
Nodes use the first slot (\ie Glossy flood) in each round also to (re)synchronize with the host.
As a result, all nodes share a common notion of time, which enables LWB's globally time-triggered operation.

\furtherExamples Several protocols and services (see \secref{sec:serv}) adopt sequential Glossy floods from LWB, but use a different scheduler and modify the structure of rounds.
For instance, in the Wireless Transparent Sensing Platform (WTSP)~\cite{suzuki17wireless}, a \emph{sink} node schedules many-to-one communication toward itself, one-to-many dissemination from itself, and one-to-one ``ping'' messages between itself and any other node in the network.
WTSP exploits the fact that one end of the communication is always the sink for end-to-end retransmissions based on implicit acknowledgments.
Blink~\cite{zimmerling17adaptive} provides real-time communication among arbitrary sets of senders and receivers.
Stream requests have an additional \emph{relative deadline} that specifies the maximum time between the generation of a packet and its reception.
The Blink scheduler allocates slots in an earliest-deadline-first manner, using an admission-control scheme to ensure that all stream requests can be served.
With this Blink guarantees that any \emph{received} packet meets its deadline.

A few many-to-many protocols depart from sequential one-to-all flooding.
For example, ByteCast~\cite{saha2017design} targets the many-to-many exchange of small data items, using an approach where senders adjust their transmit power \emph{during} a synchronous transmission so a receiver can capture data items from different senders.
Instead, Codecast~\cite{mohammad2018codecast} and Mixer~\cite{herrmann2018mixer} employ network coding to achieve a better scalability with the number of messages to be exchanged (see \secref{subsubsec:differentpackets} for details).
Although these protocols deliver every message to all nodes by default, packet filtering as in LWB can be used to only pass messages to the higher layers if the node is an intended receiver. 


\subsubsection{Many-to-one}
\label{sec:many-to-one}

Protocols in this class enable multiple source nodes in the network to transmit data to a sink node, which may provide connectivity to edge and cloud computing infrastructure.

\examplePaper{Crystal} Crystal~\cite{istomin16crystal} is a data collection protocol for aperiodic traffic.
Nodes wake up at the beginning of an \emph{epoch}, which consists of a short active phase and a long sleep phase.
The active phase starts with a Glossy flood by the sink to synchronize the nodes.
One or multiple pairs of slots follow.
The first slot of every pair is similar to a contention slot in LWB: Multiple nodes with pending traffic may concurrently initiate a Glossy flood with a \emph{different} packet.
In the second slot, the sink acknowledges the reception of one of the packets due to the capture effect---or the reception of no packet---through another Glossy flood.
Nodes whose packet is not acknowledged retransmit it in the first slot of the next pair.
The sink enters the sleep phase if it does not receive any packet for a pre-defined number of consecutive slots.
A distributed termination policy ensures that all other nodes also go to sleep after a certain number of slots with no network activity.


Crystal demonstrates that this functioning can translate into a highly energy-efficient runtime operation when combined with data prediction.
The key idea of data prediction is that the sink estimates the physical quantity being sensed using a model.
Thus, instead of periodically reporting its sensor readings, a node only sends a model update (\ie metadata) to the sink if the difference between the sensor reading and the model prediction exceeds a threshold.
Previous studies have shown that this approach can significantly reduce the number of data transmissions~\cite{raza2015practical, santini2006adaptive}.
As aperiodic model updates should be received by the sink, an enhanced version of Crystal~\cite{istomin2018interference} adds channel hopping and noise detection to increase the protocol's resilience to interference.

\furtherExamples Another synchronous transmission based protocol that targets many-to-one communication is Choco~\cite{suzuki13low}, in which the sink schedules a Glossy flood for each packet including end-to-end retransmissions.
The scheduler in Sleeping Beauty~\cite{sarkar2016sleeping} also runs at the sink.
It uses information about each node's one-hop neighbors to determine connected subsets of active nodes that transmit their packets, one after another, to the sink using Glossy-like floods, while all other nodes are put to sleep. 
The protocol proposed by Zimmerling et al.~\cite{zimmerling13modeling} directly builds upon LWB~\cite{ferrari12lwb} and assumes that the sink and the host are the same physical device.
The protocol also performs end-to-end retransmissions.
Moreover, each node piggybacks a few quantities about its runtime operation on the packets.
Feeding this information into models, the sink can provide probabilistic reliability guarantees and predict the long-term energy consumption of each node.

\subsubsection{All-to-all}
\label{subsubsec:all-to-all}

Protocols in this class enable all nodes in a network to share data with each other, forming the basis for network-wide agreement or for computing a network-wide aggregate.

\examplePaper{Chaos} The first synchronous transmission based protocol specifically designed for this purpose is Chaos~\cite{landsiedel13chaos}.
In Chaos, a designated node starts an all-to-all interaction by transmitting its message.
When a node receives a message, it combines its own data with the received data according to a \emph{merge operator}, then transmits the combined data synchronously with other nodes.
A merge operator can be, for example, a function that computes the maximum of a set of numbers or that concatenates a node's own and the received data.
To avoid unnecessary transmissions, nodes are only allowed to transmit if they have new information to share.
Whether there is new information to share is determined using \emph{flags} inside each message, where each flag is mapped to a node in the network: A node sets a flag to 1 if it has (directly on indirectly) received the data of the corresponding node by computing the XOR of its own flags and the received flags.
Once a node sees that all flags are set to 1 (\ie it has received data from all nodes), it blindly transmits the resulting message multiple times to allow other nodes to quickly reach completion.

\furtherExamples The many-to-many protocols from above also support all-to-all communication.
The main difference is that these protocols are meant for exchanging messages---they are pure communication protocols---whereas Chaos also performs in-network processing, allowing it to combine the data from all nodes into a single aggregate that is ultimately known to all nodes.


\subsection{Traffic Pattern}
\label{sec:traffic_pattern}

We next classify synchronous transmission based communication protocols according to their traffic pattern.
We thereby refer to the temporal pattern of transmitted packets \emph{inside} the network as determined by the protocol logic, which may be different from the pattern with which packets are generated by the application.
We distinguish protocols that transmit packets
\begin{itemize}
 \item in a \emph{time-triggered} fashion, including periodic and aperiodic traffic as governed, for example, by a schedule that is computed online or offline to determine when packets can be sent; 
 \item in an \emph{event-triggered} fashion, that is, packets can be sent instantaneously at any point in time with respect to the timescale of a packet's airtime (\ie initial delay of a few milliseconds). 
\end{itemize}

\subsubsection{Time-triggered}
\label{sec:time_triggered}

These protocols rely on a network-wide notion of physical or logical time to plan in advance when packet transmissions can occur.
Typically, this information is encoded in a global communication schedule, which is distributed to all nodes to drive the network operation.

\examplePaper{LaneFlood} LaneFlood~\cite{brachmann16laneflood} uses a network-wide schedule to carry data traffic to and from nodes in an aperiodic fashion.
Time in LaneFlood is divided into \emph{sessions}, and each session contains several \emph{rounds}.
Communication occurs at the end of each round in allocated \emph{communication slots}.
A node designated as the initiator always sends a message during the first slot of each session.
This message is used to synchronize the network and to let the initiator indicate whether or not it has data to transmit.
If it has, the intended destination of the communication attempt replies, resulting in a handshake similar to CXFS~\cite{carlson13cxfs}.
If it the designated initiator has no data, other nodes with data use the second slot in the session to send their communication requests.

After the handshake, nodes in LaneFlood remain active only if they belong to the forwarder set, defined as in CXFS.
However, instead of considering a fixed boundary $b$, LaneFlood lets nodes on a non-shortest path join the forwarder set according to a given probability.
To tune this probability LaneFlood exposes a protocol parameter called the \emph{slack}.
This mechanism allows LaneFlood to include fewer nodes than CXFS in the forwarder set, saving energy without sacrificing reliability.


\furtherExamples Indeed, \tabref{tab:comm-overview} shows most synchronous transmission based protocols exhibit a time-triggered traffic pattern.
While some of these protocols focus on periodic traffic, such as Blink~\cite{zimmerling17adaptive} and Sleeping Beauty~\cite{sarkar2016sleeping}, others are geared toward aperiodic traffic, such as WTSP~\cite{suzuki17wireless} and Crystal~\cite{istomin16crystal,istomin2018interference}.
Nevertheless, they all plan well in advance when end-to-end transmissions can occur, possibly without specifying the concrete sender.
The main reason is the need to minimize idle listening when using active radios that draw significant power even when there is no ongoing packet transmission or reception.

\subsubsection{Event-triggered}
\label{sec:event-triggered}

A few event-triggered protocols exist that do not require a persistent, global notion of time and enable nodes to transmit packets without any prior request or advance planning.

\examplePaper{Zippy} Zippy~\cite{sutton15zippy} provides on-demand one-to-all communication: individual packets are flooded upon an asynchronous network wake-up triggered by the sender.
To this end, Zippy addresses several challenges associated with enabling \st using low-complexity on-off keying (OOK) transmitters and always-on ultra-low power OOK receivers serving as wake-up radios.
When the network is idle, the transmitters of all nodes are off, but their wake-up receivers are still able to detect the presence of an incoming packet transmission while dissipating very little power (\ie on the order of a few microwatt).
When a node detects an event, it simply starts sending its message with a specific preamble using its OOK transmitter.
Neighboring nodes detect this preamble using their wake-up receivers, quickly turn on their transmitters, and then participate in propagating the message through the network using \st.

Zippy is suited to rapidly disseminate rare, unpredictable events.
Time-triggered protocols would waste significant amounts of energy to capture these events with low latency.
This is because the nodes would need to wake-up frequently to check whether there is an event to be disseminated.
However, since events occur only very rarely, the energy spent for these wake-ups is wasted.

\furtherExamples The Event-based Low-power Wireless Bus (eLWB)~\cite{sutton2017thedesign} supports event-triggered traffic, based on a commodity node platform with a single active radio.
When idle, eLWB schedules very short periodic rounds that contain only one \emph{event contention} slot.
When the need to communicate arises (\eg when an event is detected), one or multiple nodes send the same short, pre-defined packet in the event contention slot.
Then, eLWB immediately schedules an \emph{event round}, allowing all nodes to transmit their event and possibly request more bandwidth in future rounds.

BLITZ~\cite{sutton2017blitz, sutton2019blitz} combines the wake-up radio technique from Zippy~\cite{sutton15zippy} with Glossy-based data dissemination.
Specifically, the wake-up radio is only used to coordinate the start of a LWB-like communication round.
Compared with eLWB, BLITZ spares the overhead of periodic communication rounds, as these occur only when at least one event of interest is detected.
This approach greatly reduces the worst-case latency in communicating the occurrence of an event.

\subsection{Type of Packets}

Motivated by the physical-layer phenomena enabling \st and the associated requirements affecting protocol design (\eg in terms of timing), we distinguish protocols that
\begin{itemize}
 \item ensure that always only \emph{identical} data packets overlap in a synchronous transmission;
 \item allow for \emph{different} data packets to overlap in a synchronous transmission.
\end{itemize}

As before, we refer to the \emph{intent} of a communication protocol, not to what may accidentally happen at runtime.
Furthermore, we restrict our attention to what happens during the communication of application-level \emph{data}, as opposed to during the exchange of a protocol's control packets.

\subsubsection{Identical}

Table~\ref{tab:comm-overview} shows that most of the protocols we survey intend to transmit identical data packets.
One reason might be that the papers that made \st popular (\eg ~\cite{dutta10design,ferrari11efficient}) focused on exploiting non-destructive effects when identical packets collide.


\examplePaper{PEASST} A notable example of a protocol that transmits only identical data packets is PEASST~\cite{jeong14peasst}.
It enables point-to-point data transfers by exploiting low-power probing (LPP) to wake up the network before data communication.
LPP uses a three-way handshake to establish a connection between senders and intended receivers.
Receivers wake up periodically and send probe messages to indicate their availability to receive data.
A sender waiting for a specific receiver to wake up reacts to this probe by sending a reply that includes the address of the intended destination and forces nodes that receive it to keep their radios on.
The intended receiver finally confirms the reception of the message from the sender and data exchange can start.

The handshake between receivers and senders allows the protocol to determine which subset of nodes must stay awake for a specific sender-receiver pair to communicate, similar to CXFS~\cite{carlson13cxfs} and LaneFlood~\cite{brachmann16laneflood}.
Using this mechanism, PEASST can, in principle, enable different sender-receiver pairs to communicate concurrently, whereas each data transfer is mapped onto floods of identical data packets.
This differentiates PEASST from approaches like Mixer~\cite{mager17mixer} and Chaos~\cite{landsiedel13chaos}, which intentionally let different data packets overlap both temporally and spatially in the network.

\furtherExamples Glossy and most protocols that build on it fall into this category.
For instance, although in LWB multiple nodes may initiate a Glossy flood in the same contention slot (see~\secref{subsubsec:many-to-many}), the transmitted packets carry \emph{control} information.
By contrast, the first slot of each pair in Crystal is intended for the transmission of application-level data (see \secref{sec:many-to-one}), and thus Crystal falls into the other category.
Other noteworthy examples are protocols exploiting pipelined flooding, such as \pthree~\cite{doddavenkatappa14p3} and Pando~\cite{du15pipelines,du16pando}.
Using these protocols, different data packets co-exist \emph{spatially} in the network, but those that \emph{temporally} overlap at a receiver are always identical.
Thus, these protocols aim to meet the tight timing requirements of constructive interference (see \secref{sec:primer}).

\subsubsection{Different}
\label{subsubsec:differentpackets}

A number of very recent protocols use \st of different data packets.
These protocols are subject to the looser timing requirements of the capture effect, yet they must be cautious about the number of synchronous transmitters to meet the power requirements.
%

\examplePaper{Mixer} A protocol that lets different data packets occur in \st is Mixer~\cite{mager17mixer,herrmann2018mixer}.
In the very beginning of a Mixer round, nodes transmit only the original messages.
However, as soon as nodes receive more messages from their neighbors, they start transmit linear combinations of multiple original messages.
To do so, nodes mix packets using random linear network coding (RLNC)~\cite{ahlswede00network,ho2006rlnc}.
Since each node randomly, independently decides which packets to mix, nodes generally transmit different packets.
A successful reception of these packets is still possible because of the capture effect, as the transmissions are sufficiently well synchronized and Mixer steers the number of synchronous transmitters based on short-lived information about a node's neighborhood.
Nodes can decode the original messages from the received, mixed packets using the coding vectors embedded in the packet headers, which indicate what messages are coded together.
In this way, Mixer approaches the order-optimal scaling with the number of messages to be exchanged in a many-to-many fashion across dynamic wireless mesh networks.

\furtherExamples Codecast~\cite{mohammad2018codecast} also provides many-to-many communication, but uses Fountain coding instead of RLNC, leading to a different scaling behavior.
Nodes in Chaos~\cite{landsiedel13chaos} send uncoded messages in a slotted fashion like in Mixer or Codecast, yet combine data from different nodes according to a merge operator. 
ByteCast~\cite{saha2017design} uses a packet concatenation scheme to compact small amounts of data from multiple neighboring nodes into a single packet.

\subsection{Reliability Mechanisms}

As described in \secref{sec:primer}, the high reliability of synchronous transmission based protocols cannot be explained by \emph{sender diversity} alone.
In addition, \emph{receiver diversity} occurs whenever a transmission is overheard by more than one node, regardless of the number of senders.
Further, all protocols we survey leverage one or multiple of the following mechanisms to achieve high reliability:
\begin{itemize}
 \item local (\ie one-hop) or end-to-end packet \emph{retransmissions};
 \item \emph{channel hopping} to exploit frequency diversity;
 \item \emph{network coding} for increased resilience to packet loss at the same communication costs.
\end{itemize}


\subsubsection{Retransmissions} Despite very few exceptions, all communication protocols in \tabref{tab:comm-overview} use packet retransmissions (see \tabref{tab:comm-overview}), albeit in vastly different flavors as discussed below.
Retransmissions exploit \emph{temporal diversity}: If a transmission failed, then chances are that some time later the retransmission may not suffer from the same effect that caused the previous one to fail.

%

\examplePaper{Probabilistic reliability guarantees} Zimmerling et al.~\cite{zimmerling13modeling} explore the analytical modeling of LWB  to predict long-term energy consumption and to provide probabilistic reliability guarantees.
This is feasible because of two main reasons.
First, unlike link-based protocols, LWB's protocol logic is independent of the time-varying network state (link qualities, neighbor tables, \etc), so the network dynamics play no role in the modeling.
Second, the authors show empirically that, unlike link-based transmissions, packet losses and receptions in Glossy can be faithfully characterized by a series of independent and identically distributed (i.i.d.) Bernoulli trials.
Later confirmed by a theoretical study~\cite{Karschau2018}, this property enables simple, yet accurate analytical models.

To give probabilistic reliability guarantees, Zimmerling et al.\ extend the original LWB scheduler in a setting where the host is also the sink.
At the end of each round, the scheduler allocates missing packets another slot in the next round, up to \kmax times.
This is taken at the sources as an implicit acknowledgment that the sink has not received the corresponding packet, requiring an end-to-end retransmission.
As the Bernoulli property holds, it is easy to derive an appropriate \kmax value for all streams given a desired end-to-end reliability, as determined by application requirements~\cite{zimmerling13modeling}.

\furtherExamples A number of protocols use the same flavor of end-to-end retransmissions, including Crystal~\cite{istomin16crystal,istomin2018interference}, Choco~\cite{suzuki13low}, and WTSP~\cite{suzuki17wireless}.
These protocols and many others based on \st also benefit from local (\ie one-hop) retransmissions; for example, in Glossy~\cite{ferrari11efficient}, RedFixHop~\cite{escobar16redfixhop}, or Zippy~\cite{sutton15zippy} every node in the network \emph{opportunistically} retransmits a packet up to a certain number of times.
Local retransmissions may also be triggered by an explicit \emph{request} as used in Codecast~\cite{mohammad2018codecast}, Mixer~\cite{mager17mixer,herrmann2018mixer}, or Splash~\cite{doddavenkatappa13splash} to collect a few packets that some nodes still miss toward the end of a many-to-many or one-to-all communication phase.

%

\subsubsection{Channel hopping}
\label{subsubsec:multiple_channels}

This is a common technique to address sources of unreliability, such as interference from co-located networks and background noise~\cite{boano11jamlab}, through \emph{frequency diversity}: While one channel may be affected by such phenomena, likely other channels are less affected~\cite{raman10PIP}.

\begin{figure}[tb]
\begin{center}
\includegraphics[scale=.4, angle=-90]{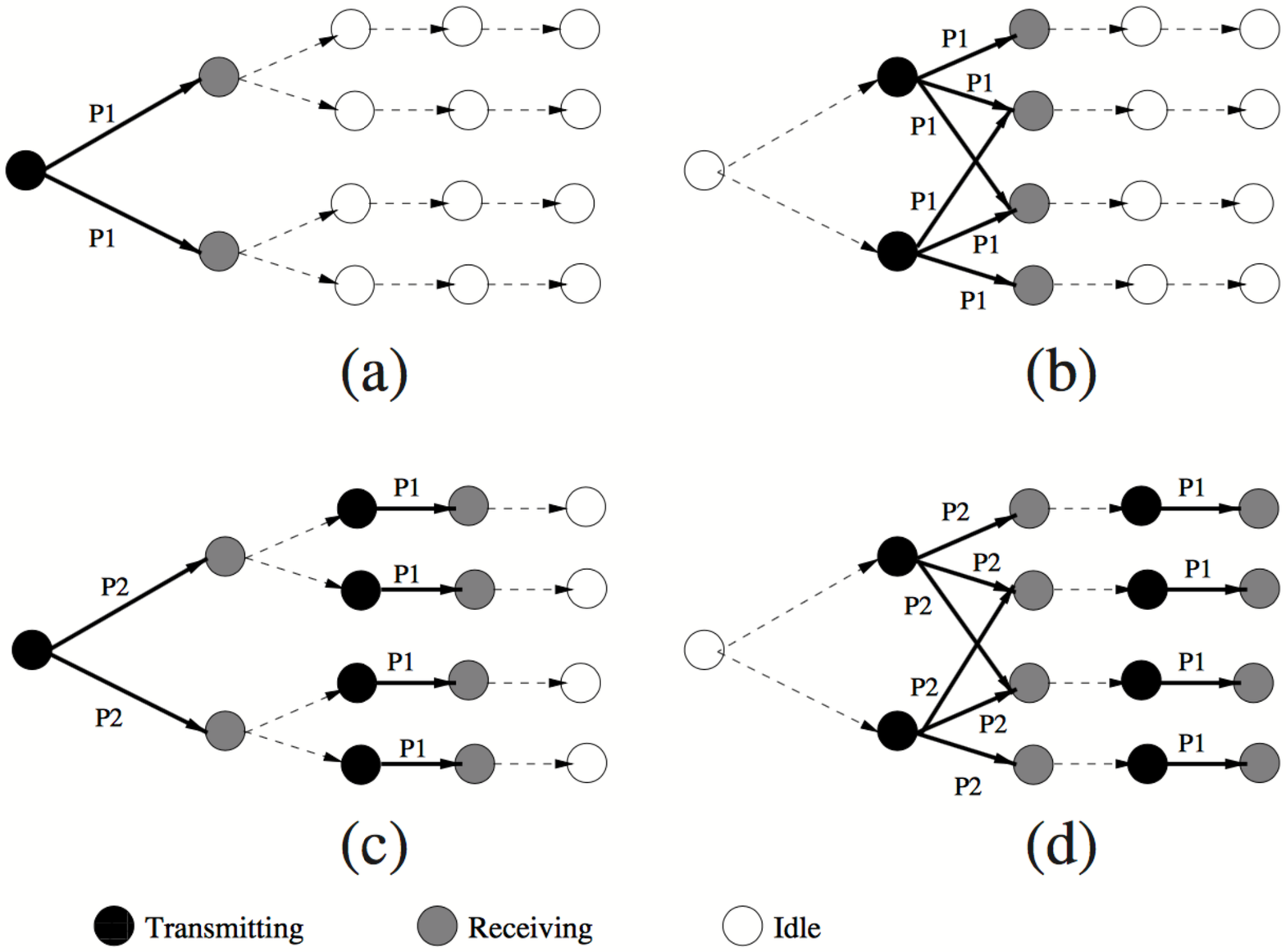}
\caption{Splash operation~(taken from \cite{doddavenkatappa13splash}). \capt{Splash creates parallel pipelines over multiple paths that cover all the nodes in the network, based on a routing tree previously established by an existing data collection protocol.}}
  \label{fig:splash}
\end{center}
\end{figure}

\examplePaper{Splash} The Splash protocol~\cite{doddavenkatappa13splash} uses pipelined Glossy floods for fast and reliable dissemination of large data objects (\eg a program image).
The source divides the data object into multiple smaller packets.
Nodes at the same hop distance from the source form a \emph{layer}, and every other layer simultaneously transmits a different packet, as shown in \figref{fig:splash}.
This way, Splash creates parallel pipelines over multiple paths that cover all the nodes in the multi-hop network.

Inspired by the pipelined transmission scheme from~\cite{raman10PIP}, membership of a node to a layer is decided based on an existing routing tree created by a traditional link-based data collection protocol (\eg~\cite{gnawali2009ctp}).
Splash assumes that such collection protocol runs between the dissemination phases, and the root of the tree is taken as the dissemination source.
A pre-determined channel assignment that maps a layer to a channel is distributed prior to each dissemination round.
Consecutive layers in the pipeline are assigned the same channel, whereas channel changes across the pipeline employ no adjacent channels.
This makes Splash reduce internal interference, support deeper pipelines, and work around interference that extends across several hops.
In addition, Splash also incorporates a scheme to reduce the number of transmitters in areas with a high node density, opportunistic overhearing of transmitted packets in the same layer, and XOR coding of packets (see \secref{sec:coding}).

\furtherExamples Several synchronous transmission based data dissemination and bulk-transfer protocols, which need to transmit multiple packets from the same source, use a similar channel-hopping approach in combination with pipelined Glossy-like flooding, such as Ripple~\cite{yuan15ripple}, Pando~\cite{du15pipelines,du16pando}, and \pthree~\cite{doddavenkatappa14p3}.
BlueFlood~\cite{alnahas2019concurrent} performs channel hopping while flooding a single packet: Every node switches to a new channel in each time slot \emph{within} a flood according to a network-wide hopping sequence, which covers all 40 channels available in Bluetooth 5.
Instead, the extended version of Crystal~\cite{istomin2018interference} changes channel from one Glossy flood to the next, but not within a flood.

\subsubsection{Network coding}
\label{sec:coding}

Rather than forwarding messages as independent units through the network, communication protocols using network coding deal with packets that are combinations of multiple messages~\cite{ahlswede00network}.
To this end, senders, receivers, and relay nodes encode and decode packets, depending on the network-coding technique used.
In general, network coding helps achieve higher throughput or higher reliability (or a middle ground) at the same communication costs compared to the non-coding case.
It has been found that network coding is particularly suitable for wireless and sensor networks, for example, due to the broadcast and lossy nature of the wireless medium~\cite{katti08xors,fragouli2006nc}.


\examplePaper{Pando} Similar to Splash~\cite{doddavenkatappa13splash}, Pando~\cite{du15pipelines, du16pando} disseminates large data objects using pipelined Glossy floods.
Instead of distributing the original messages that constitute the data object, the source injects encoded packets built by encoding a specific number of randomly selected original packets.
All other nodes relay and incrementally decode these packets as they arrive, using a variant of the Gaussian elimination algorithm.
Nodes recover the data object when they receive as many linearly-independent linear combinations as there are original messages.
Afterward, Pando uses a silence-based feedback scheme to inform the source that the dissemination process can stop.
Unlike Splash, where nodes use different channels but stay on a specific one for an entire dissemination round, nodes in Pando switch channel on a per-packet basis to better adapt to channel variations.


\furtherExamples Only a few other communication protocols exist in the literature that combine \st with different forms of network coding.
While Pando~\cite{du15pipelines, du16pando} leverages Fountain coding, Splash~\cite{doddavenkatappa13splash} uses XOR coding and Ripple~\cite{yuan15ripple} uses a Reed-Solomon code for fast and reliable one-to-all data dissemination.
Codecast~\cite{mohammad2018codecast} and Mixer~\cite{mager17mixer,herrmann2018mixer} target fast, reliable many-to-many communication, using Fountain coding and RLNC, respectively.
Both protocols exploit feedback from neighbors, enabling a sender to build encoded packets that are likely useful for its neighbors, that is, help toward a fast completion of the many-to-many exchange.

%
%

\subsection{Network State}

Whether the logic of a protocol relies on information about the state of the network is a key factor determining the protocol's ability to adapt to changes resulting from moving or failing nodes, time-varying external interference, \etc
Traditional link-based protocols maintain network state to \emph{avoid} packet collisions, for example, by scheduling transmissions between individual sender-receiver pairs based on information about a node's one-hop neighbors and the quality of the links between the node and its neighbors.
Instead, \st \emph{embrace} packet collisions, enabling the design of communication protocols that maintain no network state at all.
Nevertheless, the design space of existing protocols is diverse, motivating us to distinguish protocols whose logic
\begin{itemize}
 \item is independent of any information about the network state, referred to as \emph{none};
 \item depends on \emph{static} network state information that is not updated by the protocol;
 \item depends on \emph{dynamic} network state information that is updated by the protocol.
\end{itemize}

We note that while a protocol's \emph{logic} may be independent of the network state, its \emph{performance} always depends on the network state.
For example, the time and energy needed to transmit a packet from one node to another inevitably depends on the current hop distance between the two nodes.

%
%
%
%

\subsubsection{None}
\label{sec:no_network_state}

Protocols in this class are oblivious of the network state, allowing them to seamlessly operate in the presence of significant network and environment dynamics.

\examplePaper{BlueFlood} The work on BlueFlood~\cite{alnahas2019concurrent} explores through controlled experiments the applicability of \st to the physical layer of Bluetooth 5 and the impact of several factors, from time offsets to packet contents.
The authors then build on these insights and propose a Glossy-like one-to-all protocol that maintains no network state.
Different from Glossy, every node in BlueFlood performs a certain number of back-to-back synchronous transmissions after receiving a packet, which increases efficiency compared with Glossy, and switches channel after each transmission and reception for increased reliability, as discussed in \secref{subsubsec:multiple_channels}.
The nodes must keep track of round and slot numbers to follow the global channel hopping sequence, yet this kind of state only depends on the progression of time and not on the time-varying network state.

\furtherExamples Looking at \tabref{tab:comm-overview}, we find that many of the communication protocols we survey maintain no network state.
All of them adopt some form of synchronous transmission based one-to-all flooding, either individually to transmit a single packet~\cite{ferrari11efficient,escobar16redfixhop,sutton15zippy} or in a series to transmit multiple packets~\cite{ferrari12lwb,sutton2017thedesign,sutton2017blitz,sutton2019blitz,suzuki17wireless,zimmerling17adaptive,zimmerling13modeling,suzuki13low}.
The only exception is Chaos~\cite{landsiedel13chaos}, where nodes transmit different packets in a slotted fashion for all-to-all comunication.

\subsubsection{Static}
\label{sec:location}

A few protocols operate on network state information that is assumed to be static for extended periods of time and not updated by the protocol itself.
One of the goals of these protocols is to limit the number of nodes involved in flooding a given packet to improve efficiency.

\examplePaper{SCIF} The Spine Constructive Interference-based Flooding (SCIF) protocol~\cite{wang12exploiting, wang13exploiting} applies graph algorithms to identify a backbone to funnel packets through.
The authors observe analytically that involving all nodes in the flooding creates reliability problems as the number of synchronous senders increases.
To address this issue, SCIF determines a network \emph{spine}, a subset of the nodes showing specific topological characteristics that act as a backbone for the flooding.
Unlike in Glossy, nodes outside of the spine remain silent after receiving a packet.
This results in fewer synchronous senders, improving reliability according to the author's analytical results.

To determine the spine, SCIF assumes that node locations are \emph{known a priori} and \emph{do not change} over time.
With this information, and assuming a unit-disk graph model with known communication range, the network topology is mapped to a graph structure.
The spine is determined by dividing the resulting graph in smaller cells with a known ratio to the communication range, and by enforcing cells to form a connected sub-graph through at most one link between adjacent cells.
Any change in the physical network topology, especially due to node mobility, would break SCIF's functioning.

\furtherExamples Splash~\cite{doddavenkatappa13splash} and Pando~\cite{du15pipelines,du16pando} do tolerate network state changes, but only \emph{between} data dissemination phases, which can take a few tens of seconds or even longer for a program image.
Both protocols rely on a tree topology, maintained by \emph{another} collection protocol, which must remain intact throughout the protocol execution, as described in \secref{subsubsec:multiple_channels}.
None of the two protocols adapts the tree topology to network state changes during the data dissemination.

\subsubsection{Dynamic}
\label{sec:tree}

Protocols exist that seek to improve performance by leveraging network state, which they update dynamically to ensure that it is consistent with the physical network topology.

\examplePaper{\pthree} The Practical Packet Pipeline (\pthree) protocol~\cite{doddavenkatappa14p3} is similar to Splash, yet optimized for one-to-one bulk transfer.
Noting the same reliability problems in dense scenarios reported in earlier works~\cite{doddavenkatappa13splash, wang12exploiting}, \pthree exploits a sub-group of nodes that form a set of node-distinct paths of a given hop length, connecting source and destination.
Based on knowledge of the network topology, obtained at a base station during a discovery phase, \pthree uses depth-first search to find the largest set of available node-distinct paths of a given hop length.
If no such path exists, the procedure repeats by increasing the desired hop length until a minimum number of node-distinct paths is found.
All nodes that do not lie on such paths do not participate in the data transfexr.
The selection of node-distinct paths is then communicated to the nodes before the bulk transfer commences.
The operation of \pthree is therefore a function of how accurate is the topology information used to identify the node-distinct paths.
For example, if a node moves between the time topology information is acquired and the bulk transfer commences, the node-distinct paths are no longer guaranteed to feature the same characteristics.
As in Splash, transmissions are pipelined over multiple channels to achieve high throughput.
In addition, the receiver uses negative acknowledgements (NACKs) to notify the sender of missing packets, thus ensuring reliable transfer of bulk data.

\furtherExamples Other protocols including CXFS~\cite{carlson13cxfs}, LaneFlood~\cite{brachmann16laneflood}, and Spleeping Beauty~\cite{sarkar2016sleeping} also exploit hop counts and link qualities dynamically collected during a discovery phase.
The ability to still tolerate node mobility and other dynamics relies on the time span over which such network state must remain consistent.
For example, nodes in Mixer~\cite{mager17mixer,herrmann2018mixer} keep a short history of what happened in the last couple of slots during a many-to-many communication phase to make informed protocol decisions, yet the time span of the history (tens of milliseconds) is so short that nodes moving at a speed of 60\,km/h do not affect the protocol's performance~\cite{herrmann2018mixer}

\section{Synchronous Transmissions in Network Services}
\label{sec:serv}

The use of \st extends beyond the design and implementation of communication protocols.
The literature includes a range of network services employing \st for a variety of purposes.
We broadly categorize these works in the following classes, and provide a representative from the existing literature to make the discussion concrete.
Table~\ref{tab:services-overview} comprehensively lists the works we discuss here, along with their classification.


\addtolength{\tabcolsep}{-3pt}
\begin{table}
 \centering
 \caption{Overview of surveyed network services using \st, classified according to the kind of service they provide. Approaches marked in bold are those described as the representative example of their own service class. }
 \begin{small}
 \begin{tabular}{|l|l|l|}
  \hline
  \textbf{System}                                         & \textbf{Service class} & \textbf{Discussion}\\
  \hline
  Virtus~\cite{ferrari13virtual}                          & Consensus & \multirow{3}{*}{\secref{sec:consensus}}\\
  \cline{1-2}
  A$^2$~\cite{al2017network}     & \textbf{Consensus} & \\
  \cline{1-2}
  Paxos Made Wireless~\cite{poirot19paxos}                                & Consensus &  \\
  \hline
  DRP~\cite{jacob16end-to-end}                                    &  Time-bounded coordination & \multirow{2}{*}{\secref{sec:timebounded}}\\
  \cline{1-2}
  TTW~\cite{jacob2018ttw}                                    &  \textbf{Time-bounded coordination} & \\
  \hline
  DMH~\cite{wang12direct}                                    &  Time synchronization & \multirow{4}{*}{\secref{sec:timesynch}}\\
  \cline{1-2}
  Reverse flooding~\cite{terraneo15reverse}                                    &  \textbf{Time synchronization} & \\
  \cline{1-2}
  TATS~\cite{Lim:2016:TAT:2893711.2893732}                                    &  Time synchronization & \\
  \cline{1-2}
  PulseSync~\cite{Lenzen:2015:PES:2823437.2823440,Lenzen:2009:OCS:1644038.1644061}                                    &  Time synchronization &\\
  \hline
  SurePoint~\cite{kempke2016surepoint}                                    &  \textbf{Localization} & \multirow{3}{*}{\secref{sec:localization}}\\
  \cline{1-2}
  Chorus~\cite{Corbalan:2019:CUC:3302506.3310395}                                    &  Localization & \\
  \cline{1-2}
  SnapLoc~\cite{Stocker:2019:SUU:3302506.3312487}                                    &  Localization & \\
  \hline
  A-MAC~\cite{dutta12amac,dutta10design,dutta08backcast}                                    &  \textbf{Medium access} & \multirow{4}{*}{\secref{sec:mac}}\\
  \cline{1-2}
  NDI-MAC~\cite{liu2016non}                                    &  Medium access & \\
  \cline{1-2}
  PiP~\cite{zhang2018concurrent}                                     &  Medium access & \\
  \cline{1-2}
  Aligner~\cite{liu2019aligner}                                    &  Medium access & \\
  \hline
  TriggerCast~\cite{wang13triggercast}                                    &  Per-hop reliability & \multirow{3}{*}{\secref{sec:perhop}} \\
  \cline{1-2}
  CIRF~\cite{yu14cirf}                                    &  \textbf{Per-hop reliability} & \\
  \cline{1-2}
  Disco~\cite{wang15disco}                                    &  Per-hop reliability & \\
  \hline
  Backcast~\cite{dutta08backcast}                                    &  Collision resolution & \multirow{4}{*}{\secref{sec:collision}}\\
  \cline{1-2}
  Strawman~\cite{osterlind12strawman}                                    &  \textbf{Collision resolution} & \\
  \cline{1-2}
  Stairs~\cite{ji14walking}                                    &  Collision resolution & \\
  \cline{1-2}
  POC/POID~\cite{wu2014fast}                                    &  Collision resolution & \\
  \hline
 \end{tabular}
 \end{small}
 \label{tab:services-overview}
\end{table}


\subsection{Consensus}
\label{sec:consensus}

Many communication protocols based on \st implement a form of synchronous behavior to coordinate operations and employ network-wide flooding to distribute application information.
These features incidentally bring the system one step closer to the synchronous broadcast model that traditional distributed systems literature builds upon~\cite{birman1987exploiting}.
Functionality such as virtual synchrony~\cite{ferrari13virtual}, two- and three-phase commit~\cite{al2017network}, as well as Paxos agreement~\cite{poirot19paxos} become possible as a result.
Notably, prior to the emergence of \st, literature argued that achieving these functionality in low-power wireless networks was exceedingly difficult~\cite{stankovic2005opportunities}.

\examplePaper{A$^2$} The Agre\-ement-in-the-Air (A$^2$) protocol~\cite{al2017network} provides a means to reach network-wide consensus, for example, as implemented using two- and three-phase commit, based on \st.
It supports applications with low-latency and reliable data delivery requirements.
A$^2$ builds on a synchronous transmission kernel called Synchrotron, which extends the Chaos~\cite{landsiedel13chaos} protocol with high-precision synchronization, time-slotted operation, a network-wide scheduler, frequency hopping, multiple parallel channels, and security features.

On top of Synchrotron, A$^2$ builds a voting protocol used to implement two- and three-phase commit, functionality for consistent group membership, and
reliable primitives for nodes to join and leave the network.
Applications use A$^2$ to reliably agree on, for example, cryptographic keys, channel-hopping sequences, or actuator commands, even in the presence of node or link failures.

The authors demonstrate that A$^2$ incurs in a 475~ms latency to complete a two- phase commit over 180 nodes in the presence of unreliable links.
This comes at an energy cost that does not exceed a 0.5\% duty cycle for 1-minute intervals.
When adding controlled node failures, they also show that two-phase commit as implemented by A$^2$ ensures transaction consistency, while three-phase commit provides liveness at the cost of possible inconsistency in specific failure scenarios.

\subsection{Time-bounded Coordination}
\label{sec:timebounded}

Existing literature also aims at capitalizing on the synchronous behavior at the root of many communication protocols using \st by extending the corresponding semantics to application-level functionality~\cite{jacob16end-to-end,jacob2018ttw}.
Especially whenever real-time requirements are at stake, the ability to comprehensively cater for timing aspects across application and network, and in a distributed fashion, allows systems to operate on a single clock reference.
This enables better resource utilization and more accurate scheduling.

\examplePaper{TTW} Jacob et al.~\cite{jacob2018ttw} design a system that operates across the stack, from application down to individual packet transmissions, in an attempt to bring the operation of a low-power wireless network as close as possible to that of a wired field bus.
To this end, three issues are to be solved: \emph{i)} dedicated scheduling techniques need to be devised to match the round-based execution used in low-power wireless to reduce energy consumption, \emph{ii)} pre-computed schedules must be determined as computing schedules entirely at run-time may be prohibitive because of resource and time constraints, and \emph{iii)} adaptive mechanisms are to be created to cope with wireless dynamics and changing requirements.

Using Integer Linear Programming (ILP) techniques, Jacob et al.\ formally specify and solve the scheduling problem across distributed application tasks, messages flowing in the network, and communication rounds.
The corresponding solution guarantees application timings to be within stated application requirements, minimizes end-to-end latency across the data flows between application tasks, and ensures safety in terms of conflict-free communication, also in the presence of packet losses.
Time-Triggered-Wireless (TTW) is a system design that incorporates the solution to the scheduling problem, therefore aiming at fulfilling staple requirements of distributed applications running on low-power wireless networks, such as reliability, timing predictability, reduced application-level end-to-end latency, energy efficiency, and runtime adaptability.
The formulation and solution to the scheduling problem, as well as the TTW design, build on top of LWB~\cite{ferrari12lwb}, as they inherit its round-based operation and the use of a host node for coordination.

The evaluation is based on simulations obtained with time and energy models and aim at quantitatively measuring the minimal latency attainable across application tasks and the energy savings obtained from round-based communication.
The results indicate, for example, a minimum latency of 50~ms in a 4-hop network using 5 communication slots.
The authors expect the simulation results to be on par with the actual performance of a concrete implementation, based on the quantitative observations obtained from synthetic models of LWB~\cite{zimmerling13modeling}.

TTW has been instrumental for the first work that demonstrates fast feedback control and coordination with closed-loop stability guarantees over real multi-hop low-power wireless networks~\cite{mager2019}, provding global schedules for application and communication tasks with negligible end-to-end jitter and minimum end-to-end latency also for multi-mode CPS~\cite{baumann19fast}.

\subsection{Time Synchronization}
\label{sec:timesynch}

Leveraging the synchronization already required among synchronous senders to reap the benefits described in \secref{sec:primer}, many communication protocols based on \st already provide a form of time synchronization~\cite{ferrari11efficient,doddavenkatappa13splash}.
Works exist that take this one step further, and focus exclusively on employing \st to achieve accurate system-wide time synchronization~\cite{wang12direct,terraneo15reverse,Lim:2016:TAT:2893711.2893732,Lenzen:2015:PES:2823437.2823440,Lenzen:2009:OCS:1644038.1644061}.

\examplePaper{Reverse flooding} Terraneo et al.~\cite{terraneo15reverse} build on the observation that an accurate implementation of \st for network-wide flooding must also remove time inaccuracies at MAC layer.
This is necessary to retain the temporal alignment of synchronous transmitters, as discussed in \secref{sec:intent}.
As a result, the challenge of time synchronization moves from accurate local timestamping at MAC layer, as seen in previous literature~\cite{elson2003wireless}, to the estimation of network propagation delays, which were previously considered a secondary issue.

To tackle this issue, Terraneo et al.\ design a compensation technique to account for such delays in a setting with one clock reference.
The core of their solution lies in identifying the predecessor set of every node in the network during flooding, and then performing per-hop propagation delay measurements similar to round-trip estimations.
The resulting measurements are employed to infer a cumulated delay distribution used to estimate the propagation delays at every hop.


Experimental evidence obtained using CC2520 transceivers and Cortex M3 MCUs in a variety of configurations indicates that their network propagation delay technique allows to cancel up to 95\% of the error induced by propagation delays for a generic clock synchronization scheme.
This allows allows Glossy-like time synchronization schemes to achieve sub-microsecond clock synchronization.

\subsection{Localization}
\label{sec:localization}

Works exist that employ \st to assist the localization in space~\cite{kempke2016surepoint,Corbalan:2019:CUC:3302506.3310395,Stocker:2019:SUU:3302506.3312487}.
In these cases, \st become a stepping stone to the development of protocols coordinating the ranging operation.
The resulting time-slotted execution allows the system to localize different tags at different points in time.

\examplePaper{SurePoint} Kempke et al.~\cite{kempke2016surepoint} present a system providing decimeter-accurate localization based on the time-of-arrival information provided by ultra-wide-band (UWB) radios, improved by frequency and polarization diversity through multiple communication channels and multiple antennas.

Compared to earlier works in RF-based localization~\cite{mao2007wireless}, SurePoint takes a step further by supporting the simultaneous localization of multiple tags.
As localization of a single UWB tag requires exclusive access to the radio spectrum, the authors borrow the idea of time-triggered mutually-exclusive operation of LWB~\cite{ferrari12lwb}, described in \secref{sec:intent}.
In essence, SurePoint replicates LWB's scheduling technique over UWB radios to determine what single tag is being localized at a given point in time, similar to the way LWB grants exclusive access to a single source at a time.
Procedures for joining the network, reaching and maintaining steady-state operation, and leaving the network are also provided, similar to LWB.

The authors demonstrate that SurePoint can handle up to 10 tags simultaneously with no impact on localization accuracy, reaching steady-state operation in only 10 seconds even when all tags enter the system at the same time.
Similar to LWB, however, the authors observe that scheduling tag localization in a exclusive-access manner scales linearly with the number of tags, which may eventually cause scalability problems.
This is an effect of the bus-like scheduling that the use of \st often leads to, as observed in multiple occasions~\cite{zimmerling13modeling, ferrari13virtual}.

\subsection{Medium Access}
\label{sec:mac}

A form of \st is sometimes employed at the MAC level to achieve efficient coordination within a single hop~\cite{dutta12amac,dutta10design,liu2016non,zhang2018concurrent,liu2019aligner}.
In this setting, \st allow one to implement quick rendezvous operations in an energy-efficient way.

\begin{figure}[tb]
\begin{center}
\includegraphics[scale=.3]{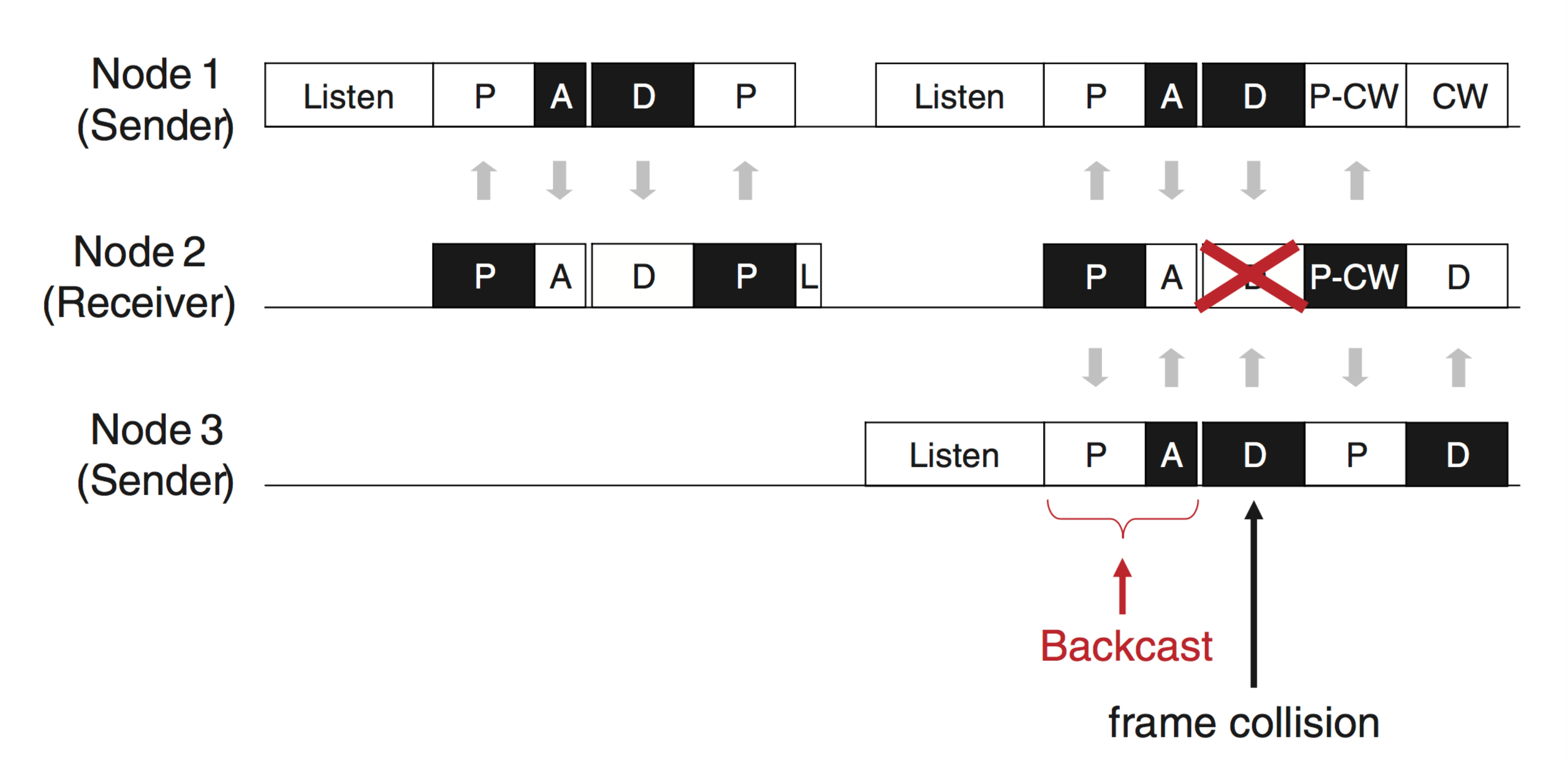}
\caption{Fundamental operation in A-MAC~(taken from \cite{dutta12amac}). \capt{The packet transmission on the left only includes a single sender. In the packet transmission on the right, the synchronous transmission of hardware auto acknowledgements allows the receiver to correctly decode the answer to the probe packet. A new contention window allows both senders to eventually transmit their packet.}}
  \label{fig:amac}
\end{center}
\end{figure}

\examplePaper{A-MAC} Dutta et al.~\cite{dutta12amac, dutta10design} present a receiver-initiated MAC protocol that, unlike previous MAC protocols of the same kind, employs a form of \st to coordinate the probing receiver with the impeding sender(s).
The basic A-MAC operation, shown in \figref{fig:amac}, requires a potential sender to first listen for a probe packet from the intended receiver, then acknowledge the frame using the 802.15.4 hardware automatic acknowledgments, then pause for a short random delay, and finally transmit the data packet if the channel is clear.

The key lies in the use of automatic hardware acknowledgements.
The 802.15.4 standard precisely dictates the content and format of such acknowledgement, as well as the time that it is generated after an application packet with the hardware acknowledgment bit set is received.
Such a time is enforced in hardware by any 802.15.4-compliant radio chip.
As a result, all nodes within the neighborhood of their intended receiver synchronously generate the acknowledgment.
Even if multiple such acknowledgments are generated, they are all identical and aligned in time; therefore, they interfere non-destructively and the receiver can decode its content with high probability.

This mechanism, called \emph{backcast}~\cite{dutta08backcast} and shown in action on the right in \figref{fig:amac}, is crucial in A-MAC, as it is the basis for the protocol to decide whether a receiver needs to stay awake as at least one impeding transmission exists, or the radio can be turned off.
Compared to how the same decision is taken in previous receiver-initiated MAC protocols, the use of \st in backcast results in quicker, more energy efficient, and more accurate operation.
The latter holds particularly in the case of multiple senders, which would otherwise generate temporal misaligned responses to the probe packet that would be impossible to decode, preventing the receiver from understanding that a transmission is about to arrive.

\subsection{Per-hop Reliability}
\label{sec:perhop}

Synchronous transmissions are also employed as a drop-in replacement for link-based transmissions in the context of protocol designs that would equally employ the latter~\cite{wang15disco,wang13triggercast,yu14cirf}.
Here, single-hop protocol interactions are adapted to make them compatible with \st without altering the traditional multi-hop protocol operation or its API. 

\examplePaper{Cirf} Yu et al.~\cite{yu14cirf} design a network-wide flooding protocol that employs a modified version of the RI-MAC protocol~\cite{Sun:2008:RRA:1460412.1460414}.
In an low-power wireless network using RI-MAC, it is likely the case that multiple nodes wait for the same neighbor to wake up when flooding.
In the original RI-MAC, contending transmitters cause collisions and a dedicated backoff mechanism is put in place to remedy this issue.
This increases packet retransmissions and latency, and is thus detrimental to energy consumption; yet, the authors observe that when these issues manifest because of network-wide flooding, the colliding packets likely carry at least the same payload.

To improve performance in this specific setting, the authors take advantage of the implicit time synchronization provided by RI-MAC's beacon packet within the one-hop neighborhood, and temporally align transmissions from multiple neighbors to leverage constructive interference, in addition to capture effect, for at least part of a packet, that is, the application payload.
A node then sleeps only whenever all its neighbors received the current packet.
This condition cannot be detected by only looking at application payloads; therefore, Cirf employs an ad-hoc packet header designed to retain the same structure across different neighbors, and thus also enjoying the benefits of constructive interference in addition to capture effect.

\subsection{Collision Resolution}
\label{sec:collision}
The literature includes works that use \st to realize efficient collision resolution mechanisms~\cite{osterlind12strawman, ji14walking, dutta08backcast, wu2014fast}.
In this case, \st are not employed to transmit actual application data, but to implement a signaling protocol invisible to upper layers.

\begin{figure}[tb]
\begin{center}
\includegraphics[width=\linewidth]{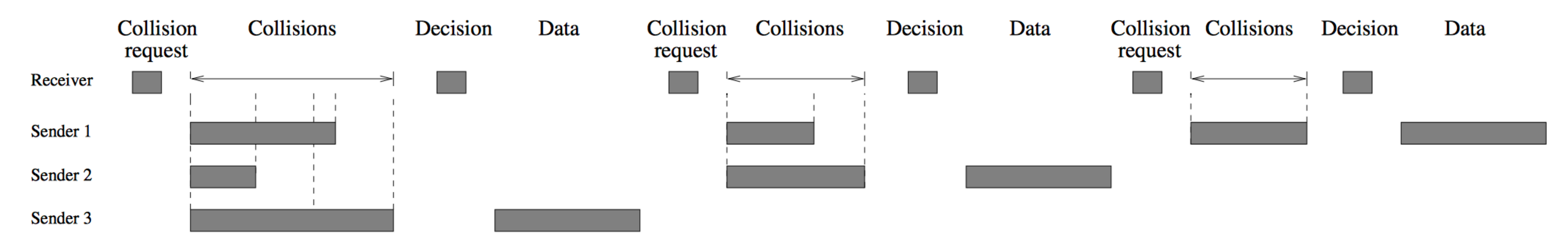}
\caption{Strawman operation~(taken from \cite{osterlind12strawman}). \capt{Upon detecting a collision, a receiver triggers synchronous transmissions of \textsc{Collision} packets from potential senders. Without concretely decoding the packet, a receiver uses the length of such packets to resolve contention.}}
  \label{fig:strawman}
\end{center}
\end{figure}

\examplePaper{Strawman} \"Osterlind et al.~\cite{osterlind12strawman} present a collision resolution mechanism for receiver-initiated MAC protocols.
\figref{fig:strawman} illustrates its operation.

Upon detecting a data collision, the receiver sends a \textsc{Collision request} packet.
The senders interpret this packet as the beginning of a Strawman round, and contend for the channel by \emph{synchronously} transmitting a \textsc{Collision} packet of \emph{random} length.
The receiver does \emph{not} attempt to decode the packet; it only needs to estimate the length of the longest \textsc{Collision} packet by sampling the received signal strength.
The receiver then broadcasts a \textsc{Decision} packet containing the longest measured length, implicitly informing the corresponding transmitter that it is now granted access to the channel.

While the selected transmitter transfers the \textsc{Data} packet, the other contenders remain silent, as they also recognized not to be given channel access based on the information in the \textsc{Decision} packet.
The subsequent \textsc{Collision request} broadcast by the receiver both initiates a new Strawman round and acknowledges the previous \textsc{Data} packet.
In the case that several contenders have chosen the same random length and their \textsc{Data} packets have collided, the receiver sends another \textsc{Collision request} packet to initiate a new Strawman round.
This process repeats until all contenders have successfully sent their \textsc{Data} packets.

In Strawman, it is therefore merely the temporal alignment of \textsc{Collision} packets that ultimately allows the synchronously transmitted packet to carry some information, without the receiver concretely decoding the packets.
Similar to A-MAC, the impact on the system's performance materializes as quicker and more energy-efficient operation compared to alternative approaches for contention resolution~\cite{jamieson2006sift}.



\section{Research Agenda}
\label{sec:agenda}

Despite the existing body of work, the features of \st and their impact on the design of communication protocols and network services constantly bear new research directions.
We make here a brief account of those, eliciting open problems and standing issues.

\subsection{Network Stack and Operation}

The design of a network stack based on synchronous transmissions is largely yet to be explored.
The protocols and services in this article represent individual prototypes, but research to consolidate and integrate those efforts into a modular, full-fledged stack that exploits the characteristics of \st with a well-defined API is limited.
There is likely no one-size-fits-all solution.
However, to best serve emerging \cps and \iot applications, research should aim for a dependable, secure, and adaptive stack that supports dynamic networks and provides provable guarantees on performance and functional properties.
Such stack currently does not exist; however, it is a challenging, yet realistic goal given the recent achievements in the field.    

A notable effort in this direction is Baloo, a framework for the design of low-power wireless stacks based on synchronous transmissions~\cite{jacob2019baloo}.
At the core of Baloo is a middleware layer that allows one to compose multiple \st primitives, while hiding low-level interrupt, timer, and radio control from protocol designers.
An additional question, however, arises:

\question{What are the trade-offs between performance, functionality, and the number of \st primitives underneath?}

Exploration in this area may be pushed as far as providing networking, localization, and time synchronization using a single physical layer and \st primitive, such as one-to-all flooding over ultra-wideband radios~\cite{kempke2016surepoint,vecchia2019playing}.
If multiple primitives are used, the increase in complexity may not necessarily justify the additional flexibility or performance improvements.

Most multi-hop protocols using \st also operate in a slotted fashion.
The network-wide slot grid emerges as the transmissions of neighboring devices must be aligned in time.
Protocols that rely on capture effect typically use explicit software routines to establish and maintain the slot grid~\cite{herrmann2018mixer,mohammad2018codecast}, while those exploiting constructive interference exclusively rely on hardware interrupts~\cite{ferrari11efficient,du15pipelines}.

The network-wide slot grid serves as a global, shared timeline for scheduling the distributed protocol operation: In every slot, each node transmits, listens/receives, or turns off the radio.
Despite the key role, mechanisms to establish and maintain the slot grid are often best effort and poorly understood.
Although the nominal slot length is known to each node, the nodes must keep the slot grid in a fully distributed fashion, that is, with no central reference clock, despite varying clock offsets across nodes.
This setting is different from most time synchronization protocols, where all nodes synchronize with a specified reference node.
Further issues are thus worth exploring:

\question{Is it possible to design and formally analyze a mechanism that guarantees all nodes stick to the slot grid within a certain tolerance range, despite network dynamics and as the number of nodes, number of slots, or topological network features vary?}
  
The problem resonates with recent work on self-organizing pulse-coupled oscillator synchronization~\cite{klinglmayr17convergence} and synchronization in networks of phase-locked loops~\cite{pollakis14synchronization}.
Therefore, we ask:

\question{Are practical mechanisms based on those theories feasible, and can such an approach enable formal guarantees on convergence and synchronization accuracy?}

A carefully-designed slotting mechanism is a prerequisite for building a dependable and efficient network stack, and may elicit important bounds on, for example, network diameter and slot length.

\subsection{Networking Abstractions}

Wireless communications are, by their very nature, broadcast: The information carried by the radio waves may be received by all devices within communication range.
However, the vast majority of today's networks, from (low-power) wireless local area networks to the Internet, is based on point-to-point interactions.
Multicast and broadcast communication are only used for specific tasks, such as device discovery and data replication.

By contrast, \st break away from the point-to-point abstraction.
Several solutions in \secref{sec:comm} adopt a form of broadcast-only communication model: every message may be received by all other nodes in the (multi-hop) network.
As discussed in \secref{sec:consensus}, the combination of this and network-wide time synchronization narrows the gap between low-power wireless systems and traditional distributed ones, where a large body of exiting literature applies~\cite{birman1987exploiting}.  
This enabled network-wide agreement~\cite{al2017network} and consensus~\cite{poirot19paxos} as well as hard real-time~\cite{zimmerling17adaptive}, virtual-synchrony~\cite{ferrari13virtual}, and closed-loop stability guarantees~\cite{mager2019}.
These are desirable functionality in the target application, which appeared out of reach in low-power wireless given their dynamic nature and resource constraints~\cite{stankovic03realtime}.
This poses a key question:

\question{Is the broadcast model a better abstraction for future low-power wireless networking using \st, or what other abstractions may reap the most benefits from \st?}

We argue that future research should reconsider this aspect in light of broadcast-only communication and network-wide synchronization.
For example, Tschudin~\cite{tschudin19broadcast} points out  the tight relationship between the broadcast model and replicated append-only logs: a data structure where every node keeps a full log of all packets ever sent by a given source.
In an analogy with interpreting communication between two processes over a pipe as appending to a FIFO buffer, replicated append-only logs may allow to ``... replace networking with arbitrary data packets by networking with coherent data structures''~\cite{tschudin19broadcast}.
Barring memory limitations on the target platform, implementing replicated append-only logs appears feasible using \st, by relying on the property that each node may receive every message.
This may pave the way to the principled design and efficient implementation of a fully decentralized network stack that does not require central entities and is fault-tolerant under non-Byzantine conditions, including message losses, network partitions, and node failures.

Further examples are distributed shared memory~\cite{coulouris11distributed} and the synchronous semantics underlying languages such as Lustre and Giotto~\cite{beneviste03synchronous}, which may raise the level of abstraction for developers and allow them to reason about timing properties.
The networking and system-level mechanisms required to efficiently implement those semantics should inform the design of the network stack and communication protocols.
Putting those mechanisms in place would equate to providing access to a large body of prior work in distributed, embedded, and real-time systems that hitherto remained applicable only to conventional networks, enabling a better understanding and even formal verification of low-power wireless systems.
Existing work shows that synchronous transmissions simplify accurate protocol modeling~\cite{zimmerling13modeling}, which is often needed for verification.

On the other hand, broadcast-only communication and provisioning for the abstractions above may not be suitable for all applications, and scalability in terms of number of devices and traffic load may be a concern.
The other side of the coin prompts us to ask:

\question{What are the application facets that make would such a decentralized approach appropriate, given the achievable consistency model, performance, and timing guarantees may be different compared with a centralized approach?}

A better understanding of the conceptual and practical trade-offs between broadcast-only and point-to-point interactions, along with their breakeven point in real networks, if any, may be informed by recent theoretical results~\cite{haeupler14broadcast,haeupler16analyzing}.

\subsection{Implications of Physical-layer Features on Protocol Design}

The works in this article indicate some of the potential of using \st instead of (or alongside with) \lt to design protocols with unprecedented functionality, performance, and efficiency.
As the community broadens its understanding of lower-level questions, such as why and under what conditions \st work for an increasing number of low-power wireless physical layers and receiver architectures~\cite{firstpart}, the corresponding insights may have profound implications on the design of higher-level communication protocols and network services.
The problem is multi-faceted.

\fakepar{Physical-layer phenomena} It is important to examine how low-power wireless receivers benefit from constructive interference versus different phenomena, such as capture effect.
Orthogonal to this, the effects of spatio-temporal diversity also warrant further investigations.

Based on current literature, we know that as the transmit bitrate of the physical layer increases, it becomes more and more difficult to meet the timing condition of constructive interference.
For example, using \wifi as physical layer, the incoming signals at a receiver must be aligned to within a few nanoseconds, that is, a fraction of the symbol length.
To meet such stringent timing constraints, special hardware and accurate network state information are a necessity, for example, to compensate for different propagation delays~\cite{rahul2010sourcesync}.
A key question therefore arises:

\question{Despite the added complexity, is it possible to employ constructive interference in networks with mobile entities or in deep networks with a diameter of more than a few hops?}

Even for low-rate physical layers, such as \ieee, it is unclear whether and to what extent the timing conditions of constructive interference are met beyond the second hop, for example, during a Glossy flood.
Sophisticated experimental setups are needed to investigate this question as currently available testbeds such as FlockLab~\cite{lim2013flocklab} do not typically provide the required synchronization accuracy.
We conjecture that ideas from time-of-flight-aware time synchronization~\cite{Lim:2016:TAT:2893711.2893732} may help increase the chances that nodes benefit from constructive interference.

Nonetheless, several arguments suggest to design protocols that exclusively build upon the capture effect.
Compared with constructive interference, the capture effect is more general and has significantly looser timing conditions, corresponding to the length of the packet preamble.  
This simplifies protocol design and implementation, especially in light of emerging hardware platforms with higher clock rates, and also leave headroom for incorporating compute-intensive security features, such as packet encryption and authentication~\cite{al2017network}, network coding operations~\cite{herrmann2018mixer,mohammad2018codecast}, and other in-network processing functions~\cite{landsiedel13chaos}.
Although further studies on capture thresholds and windows are needed~\cite{firstpart}, it also becomes easier to port a protocol to other platforms, to support physical layers with a higher transmit bitrate, and to run a protocol in a heterogeneous network consisting of nodes with different clock rates.
About capture effect, however, we ask:

\question{If capture is the dominating or even the only effect that enables packet reception beyond a certain hop distance, what does this mean for protocol scalability and reliability, particularly in dense networks?}

On the contrary, existing multi-hop protocols that are exclusively based on the capture effect typically use some form of probabilistic decision making, for example, whether a node should transmit or stay silent, to increase the chances that the capture effect occurs~\cite{landsiedel13chaos,herrmann2018mixer}.
This makes their timing behavior non-deterministic and thus harder to predict accurately compared with other solutions that also build upon constructive interference~\cite{zimmerling17adaptive,zimmerling13modeling}. 
Further research into understanding and (ideally) bounding the running time of these protocols is needed, for example, to assess their applicability in hard real-time applications.

\fakepar{Packet-level reliability}
To achieve a desired reliability, many \st schemes opportunistically retransmit the same packet~\cite{ferrari11efficient,doddavenkatappa13splash} or use end-to-end retransmissions~\cite{ferrari13virtual,zimmerling13modeling}.
Recent proposals exploit different forms of network coding for higher reliability and efficiency~\cite{mohammad2018codecast,du16pando,herrmann2018mixer}.
These techniques operate above the packet level, considering a packet an indivisible unit.
Instead, mechanisms such as forward error correction (FEC), interleaving, and exploiting partially correct packets operate at a lower level and are largely unexplored in \st.
A standing issue is therefore:

\question{What packet-level reliability techniques may complement existing solutions above the packet level when using \st?}

The concrete application of these mechanisms should be informed by physical-layer observations, such as the distribution and pattern of symbol errors in received packets when using synchronous transmissions~\cite{firstpart}.
Even though further research is necessary in this area too, we conjecture two possible trends.
First, carrier frequency offsets among synchronous transmitters cause a change in the envelope of the received signal, because the receiver cannot simultaneously compensate against all transmitters.
Depending on the magnitude of carrier frequency offsets, these envelope changes may translate into characteristic patterns in the sequence of correct and incorrect symbols.
Second, one might probably observe a sloping pattern, where symbols toward the end of a packet are more likely to be corrupted than symbols at the beginning of the packet.

If the former trend is confirmed, a protocol could combine packets based on the received signal strength (RSSI): Portions of packets received with high RSSI are likely correct and could be combined to reconstruct the complete, correct packet.
This approach exploits the fact that although the error patterns in two corrupt packets may be similar, the patterns are likely shifted so symbols that are corrupt in one packet are likely correct in the other packet, and vice versa.

Further, if the second hypothesis holds, a protocol could let a transmitter alternate between sending the payload bytes in forward and reverse order, so receivers can combine the (likely correct) initial portion of the received packets to reconstruct the complete, correct packet.
Partial packet recovery~\cite{jamieson2007partial} and packet combining~\cite{dubois-ferriere2005combining} are studied in wireless networks, yet they are unexplored in \st, where nodes typically have multiple opportunities to receive the same packet. 
When using FEC, a protocol could statically or dynamically adjust the size of the code blocks and the amount of added redundancy based on the observed distribution and pattern of symbol errors.
FEC would also represent a particularly good fit for \st schemes, where a sender often has no feedback on the successful reception of a packet by a particular node.
State-of-the-art 32-bit platforms offer ample compute power to explore the integration of block or convolutional coding into the time-sensitive operation of protocols using \st~\cite{park14improving}. 
  
\fakepar{Interpreting interfering signals}
Protocols and services in this article are specifically designed for physical layers that map the interfering signals at a receiver to \emph{at most} one packet.
However, other interpretations of interfering signals are found in the literature, and could be exploited to enhance the capabilities and improve the performance of protocols and services using \st. 
For example, by interpreting the length and magnitude of the received signal strength, synchronous transmissions have been used for single-hop agreement~\cite{boano2012jag} and counting~\cite{zeng2013counting}.
Others have looked at disentangling the interfering signals in order to decode \emph{multiple} packets, based on IEEE 802.15.4~\cite{kong2015mzig,ismail2017rnr}, LoRa~\cite{eletreby2017empowering}, and IEEE 802.11~\cite{gollakota2008zigzag}.
While received signal strength interpretations work on commodity low-power wireless platforms, existing multi-packet decoding schemes require special hardware and additional computational power, for example, a software-defined radio or a wall-powered gateway.
An open research question is thus:

\question{How to best leverage and integrate alternative interpretations of interfering signals into a communication protocol, and what are the end-to-end improvements at scale, for example, in terms of throughput and overall energy costs?}

\noindent Recent work on low-power software-defined radios~\cite{hessar2020hessar} could support this kind of efforts.

\subsection{Integration and Standardization}

Low-power wireless networks must often integrate with other systems, such as the cloud, through the Internet.
One way to fulfill this requirement this is to push characteristic features of the Internet protocol suite, like asynchronous point-to-point interactions, down into low-power wireless networks by means of dedicated designs~\cite{ko11connecting}.
Synchronous transmissions prompt a different approach: message exchanges are synchronous and point-to-point interactions are inherently discouraged, as communication tends to be broadcast.

Besides a few works focusing on end-to-end real-time guarantees in \cps applications~\cite{jacob16end-to-end,jacob2018ttw}, the interfaces required for a seamless integration are largely unexplored.
Therefore, we ask:

\question{How are network functionality of a low-power wireless network using \st exposed to the outside world and accessed at runtime?}

\noindent and also:

\question{How can a globally distributed application specify end-to-end requirements, for example, in reliability and latency, between nodes residing in different low-power wireless networks using \st?}

A stepping stone to solve this issue may be to create a form of decoupling between the required functionality.
Existing literature includes a few attempts in this direction; for example, Bolt is dual-processor platform that allows one to isolate the time-sensitive operation of \st from the (a)synchronous operation of applications or other networks and systems~\cite{sutton15bolt}.
One of the processors runs the network stack using \st, while the other processor implements the application logic.

The question, however, is foremost relevant at the higher levels of the stack, where protocols such as MQTT or CoAP operate.
Key assumptions in their designs is that multicast or broadcast communication is costly and that nodes operate asynchronously.
A variant of MQTT, called MQTT-SN, accounts for the limited resources and duty-cycled operation of battery-powered devices~\cite{mqttsn}.
On the contrary, \st enable efficient reliable broadcast communication, and requires no knowledge of per-node sleep schedules at an MQTT-SN gateway~\cite{mqttsn}.
Synchronous transmissions may rather be used as building blocks for network stacks that are conceptually closer to field busses, such as CAN or FlexRay.
A standing integration issue is therefore:

\question{What interfaces and network functionality are most appropriate or required to integrate with real-time networking systems, for example, with Time-Triggered Ethernet or field busses such as CAN or FlexRay?}

We believe that work on integration is essential for widespread adoption and standardization.
Close collaboration with industry will be key to identify the relevant platforms, benchmarking scenarios, and standardization practices.
Ideally, the identified scenarios would also serve as demonstrators, showing either functionality or performance previously unattained with existing link-based transmissions.
We argue that industrial control and autonomous robotics are particularly suitable areas for initial standardization efforts, which would greatly benefit from the end-to-end guarantees and efficient performance in mobile settings enabled by \st~\cite{zimmerling17adaptive,mager2019,ferrari13virtual}.

\section{Concluding Remarks}
\label{sec:end}

This article surveys communication protocols and network services for low-power wireless systems that exploit the concept of \st.
As \st on commodity off-the-shelf devices and existing low-power wireless physical layers become popular just over the past eight years, the field is relatively young, and yet it already witnessed large amounts of work.
This article compares the diverse communication protocol designs in a structured way and summarizes the investigated types of network services.
Together with our complementary article~\cite{firstpart}, which provides a tutorial-style introduction to the principles of \st and surveys existing research that aims to better understand or to improve the reliability of the basic technique, we allow newcomers to the field of \st to quickly obtained an overview of the foundations and of the current research art.
Although the current state of the art successfully addresses several key requirements of \cps and \iot applications, including dependability, adaptability, and efficiency, further efforts are also required in several directions, ranging from understanding the implications of physical-layer features on protocol and service designs, to addressing integration and standardization issues.

\begin{acks}
We thank Carlo Alberto Boano, Carsten Herrmann, Fabian Mager, Johannes Richter, and Romain Jacob for their constructive feedback on earlier drafts of this paper.
This work was supported in part by the German Research Foundation (DFG) within the Emmy Noether project NextIoT (grant ZI 1635/2-1).
\end{acks}

\bibliographystyle{ACM-Reference-Format}
\bibliography{biblio}

\end{document}